\documentclass[preprint,3p]{arXiv}
\usepackage[caption=false]{subfig}
\usepackage{algorithm}
\usepackage{algpseudocode}
\usepackage{amsmath}
\usepackage{url}

\usepackage{amssymb}
\usepackage{amsthm}
\theoremstyle{definition}
\newtheorem{dfn}{Definition}

\newcommand{\OT}{{\rm OT}}
\newcommand{\ICT}{{\rm ICT}}
\newcommand{\OST}{\rm OST}
\newcommand{\IST}{\rm IST}

\begin{document}
\begin{frontmatter}

\title{Polygonal Sequence-driven Triangulation Validator: An Incremental Approach to 2D Triangulation Verification}
\author[inst1]{Sora Sawai}
\affiliation[inst1]{organization={Graduate School of Fundamental Science and Engineering, Waseda University},
            addressline={3-4-1 Okubo}, 
            city={Shinjuku-ku},
            postcode={169-8555}, 
            state={Tokyo},
            country={Japan}}
\author[inst2]{Kazuaki Tanaka\footnote{Corresponding author at: Global Center for Science and Engineering, Waseda University, 3-4-1 Okubo, Shinjuku-ku, 169-8555, Tokyo, Japan.}\footnote{E-mail: tanaka@ims.sci.waseda.ac.jp}}
\affiliation[inst2]{organization={Global Center for Science and Engineering, Waseda University},
            addressline={3-4-1 Okubo}, 
            city={Shinjuku-ku},
            postcode={169-8555}, 
            state={Tokyo},
            country={Japan}}
\author[inst3]{Katsuhisa Ozaki}
\affiliation[inst3]{organization={Department of Mathematical Sciences, Shibaura Institute of Technology},
            addressline={307 Fukasaku}, 
            city={Minuma-ku, Saitama},
            postcode={337-8570}, 
            state={Saitama},
            country={Japan}}
\author[inst4]{Shin'ichi Oishi}
\affiliation[inst4]{organization={Faculty of Science and Engineering, Waseda University},
            addressline={3-4-1 Okubo}, 
            city={Shinjuku-ku},
            postcode={169-8555}, 
            state={Tokyo},
            country={Japan}}

\begin{abstract}
Two-dimensional Delaunay triangulation is a fundamental aspect of computational geometry. This paper presents a novel algorithm that is specifically designed to ensure the correctness of 2D Delaunay triangulation, namely the Polygonal Sequence-driven Triangulation Validator (PSTV). 
Our research highlights the paramount importance of proper triangulation and the often overlooked, yet profound, impact of rounding errors in numerical computations on the precision of triangulation. 
The primary objective of the PSTV algorithm is to identify these computational errors and ensure the accuracy of the triangulation output. 
In addition to validating the correctness of triangulation, this study underscores the significance of the Delaunay property for the quality of finite element methods. 
Effective strategies are proposed to verify this property for a triangulation and correct it when necessary. 
While acknowledging the difficulty of rectifying complex triangulation errors such as overlapping triangles, these strategies provide valuable insights on identifying the locations of these errors and remedying them. The unique feature of the PSTV algorithm lies in its adoption of floating-point filters in place of interval arithmetic, striking an effective balance between computational efficiency and precision. This research sets a vital precedent for error reduction and precision enhancement in computational geometry.
\end{abstract}
\end{frontmatter}

\section{Introduction}
\begin{figure}[H]
    \centering
    \includegraphics[keepaspectratio, scale=0.15]{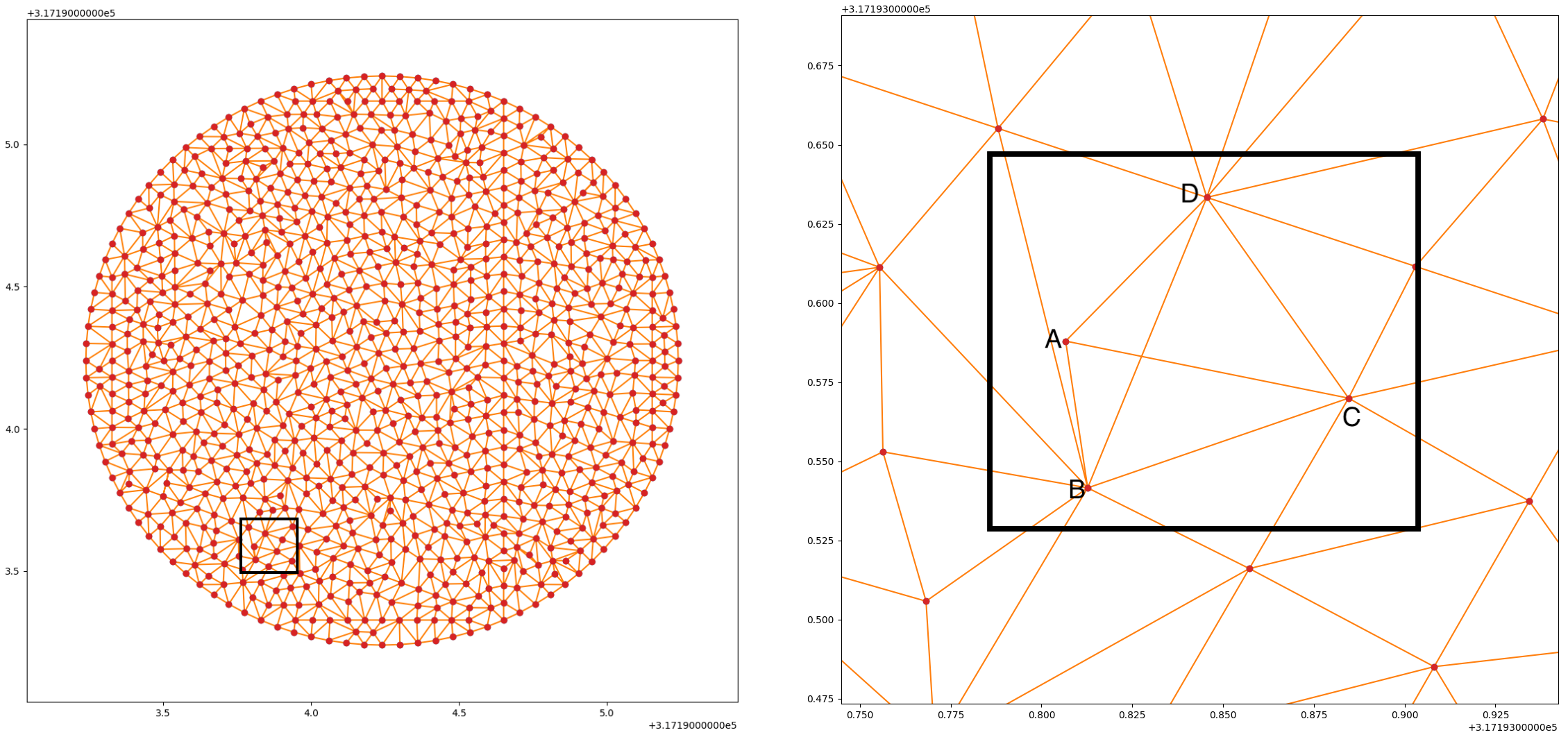}
    \caption{Triangulation (left) is obtained from the built-in `delaunay' function in SciPy. The highlighted part (right) shows overlapping triangles. The failure of the triangulation process involves overlapping, which results in shapes that are not even triangles. In relation to $\triangle{BCD}$, $\triangle{ABC}$ and $\triangle{ACD}$ overlap. The versions of SciPy and Python used for this triangulation are 1.11.0 and 3.9, respectively.}
    \label{fig:wrong-tri-1}
\end{figure}

\begin{figure}[H]
    \centering
    \begin{tabular}{cc}
        \begin{minipage}[t]{0.45\hsize}
            \centering
            \includegraphics[keepaspectratio, scale=0.15]{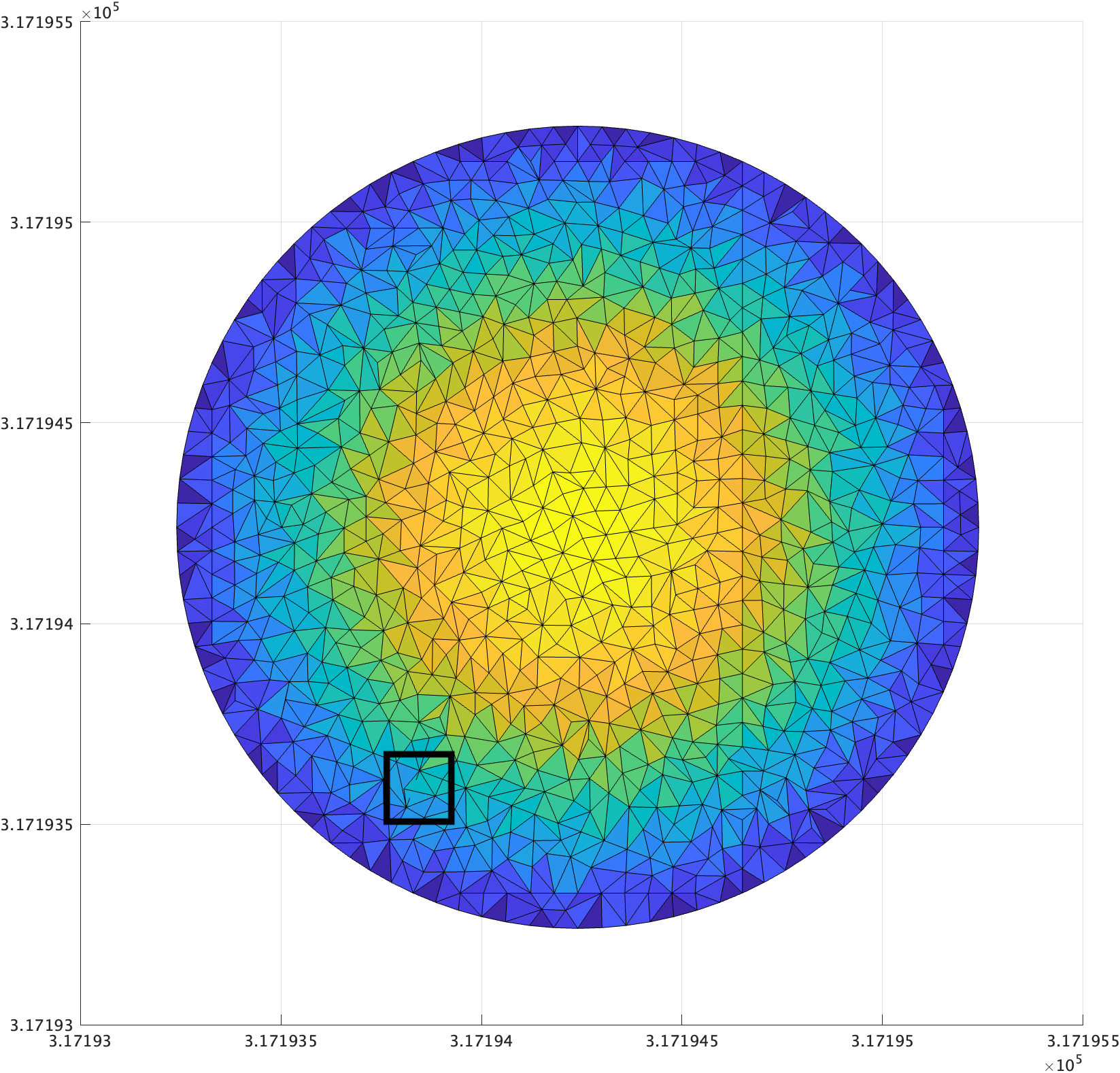}
        \end{minipage}
        \begin{minipage}[t]{0.45\hsize}
            \centering
            \includegraphics[keepaspectratio, scale=0.15]{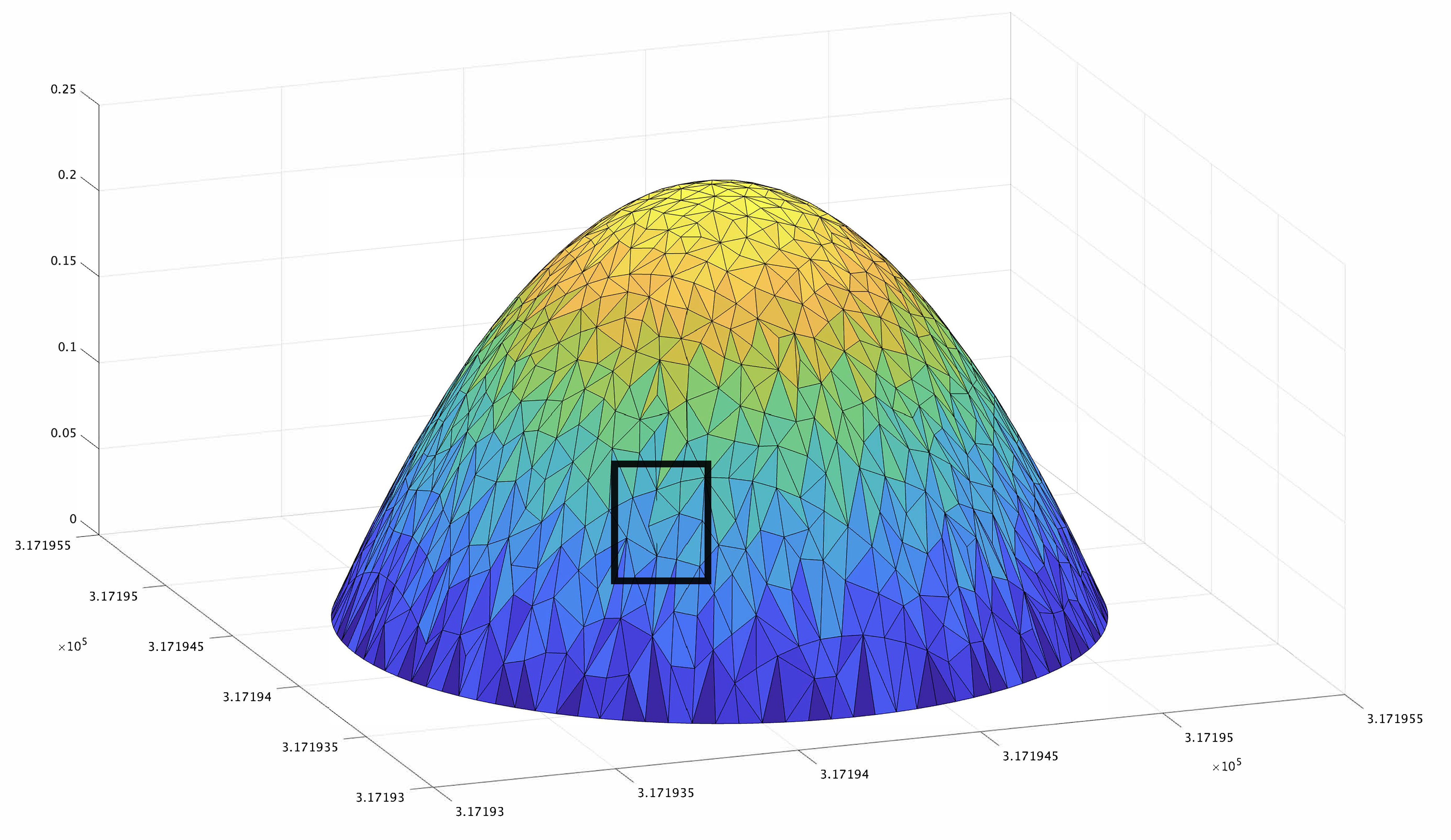}
        \end{minipage}\\\\
        \begin{minipage}[t]{0.45\hsize}
            \centering
            \includegraphics[keepaspectratio, scale=0.12]{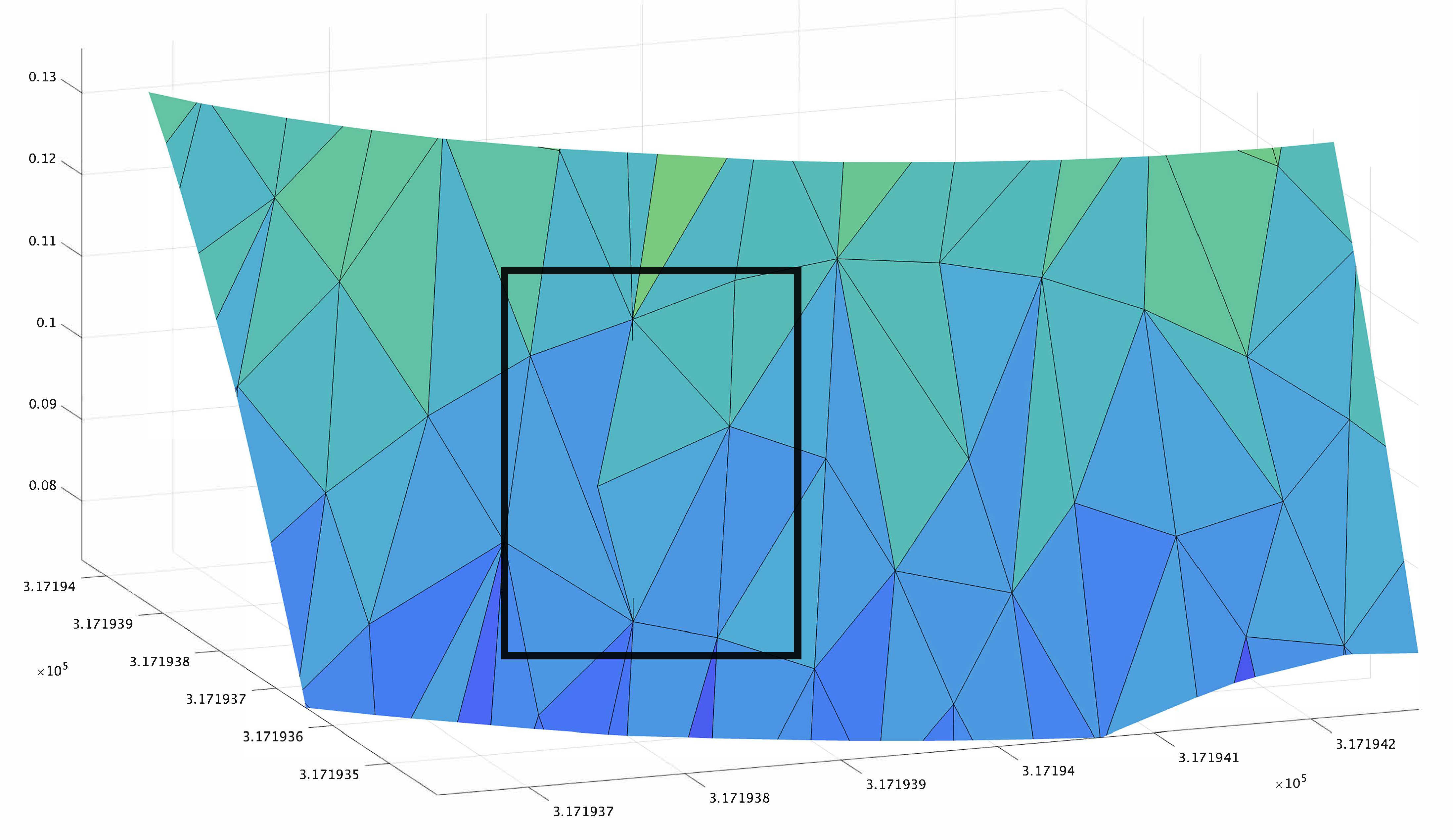}
        \end{minipage}
        \begin{minipage}[t]{0.45\hsize}
            \centering
            \includegraphics[keepaspectratio, scale=0.12]{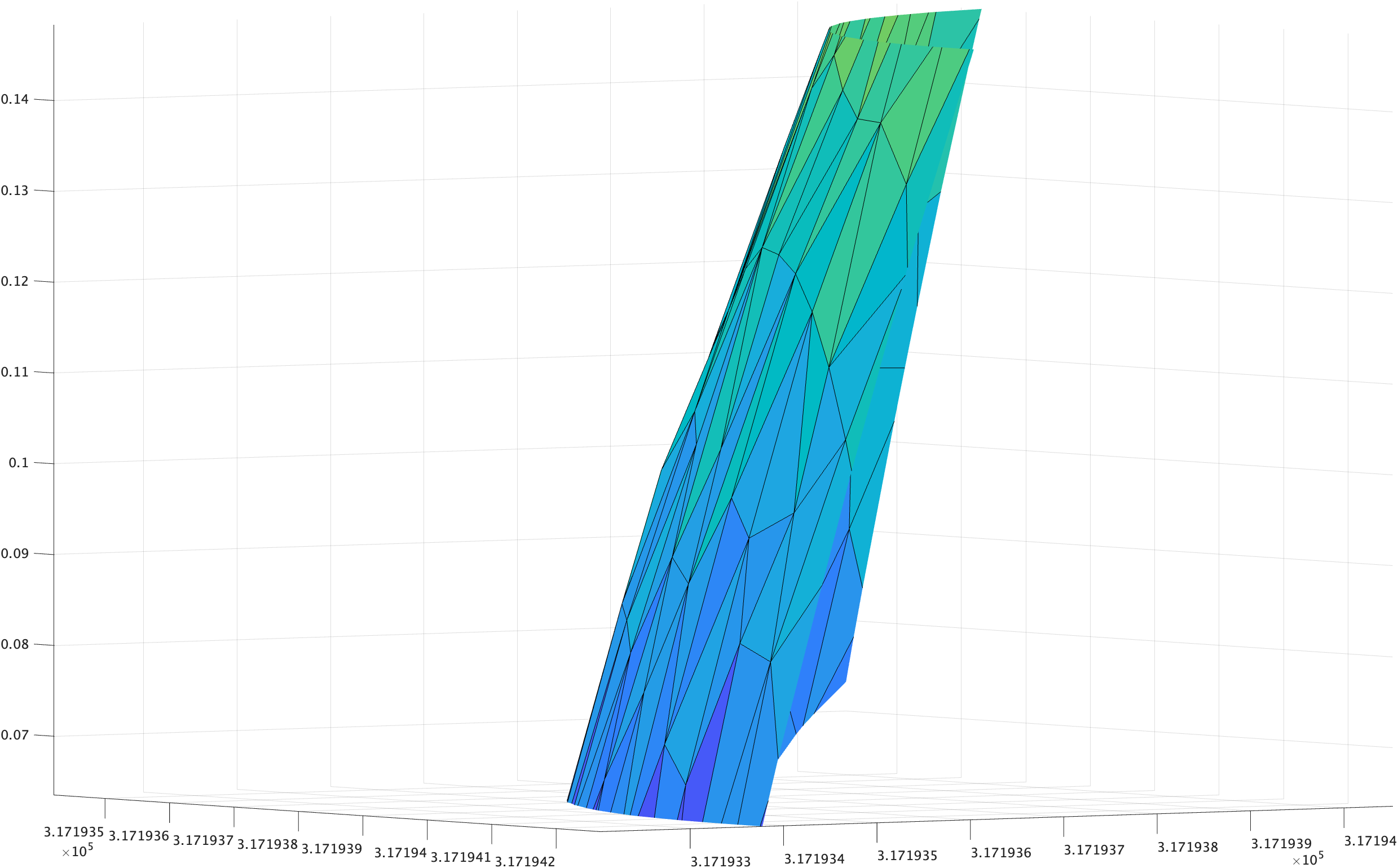}
        \end{minipage}
    \end{tabular}
    \caption{Finite element solution of the Poisson equation solved on the incorrect triangulation depicted in Fig.~\ref{fig:wrong-tri-1}. The corresponding matrix equations were solved and the visualization was produced using MATLAB R2022b.}
    \label{fig:my_label}
\end{figure}

The 2D Delaunay triangulation is a crucial component of computational geometry. It is extensively used in geographic information systems and numerical simulations of partial differential equations. The numerical computations frequently employed for generating 2D Delaunay triangulations are swiftly executed on modern computers using floating-point arithmetic. However, the finite precision of floating-point arithmetic means that rounding errors occasionally cause significant problems. Even if an algorithm is accurately developed, it may yield imprecise results when numerical computations are used.
Essential predicates are inherent in computational geometry algorithms. For instance, the 2D orientation problem entails determining whether a point lies on a line or to its left or right, whereas the incircle problem ascertains whether a point is inside, outside, or on a circle. Such problems are distilled down to the sign of a small-dimensional matrix determinant. If rounding errors occur during the evaluation of this determinant, an incorrect sign may be obtained, leading to unforeseen results. For example, the computed result of a convex hull may exclude a point, or the result might not be convex. The issues resulting from rounding errors are referred to as robustness problems; they are thoroughly summarized in \cite{kettner2008classroom}.
Figure~\ref{fig:wrong-tri-1} illustrates an incorrect triangulation computed by the `delaunay' function in the SciPy Python library.
This error is highly likely to stem from an issue related to 
finite-precision floating-point computation.
Although this is a particular situation in which the given point cloud has a large absolute value, it demonstrates that grave errors can transpire under any circumstances.
This problem potentially becomes more pronounced in applications such as the finite element method (FEM) with adaptive mesh refinement (AMR), where triangulation may include extremely small triangles, significantly increasing the likelihood of rounding errors affecting the results.
Remarkably, the Poisson equation, a fundamental partial differential equation, can be solved flawlessly on this incorrect triangulation using FEM (Fig.~\ref{fig:my_label}).
This underscores the difficulty of detecting such issues. Furthermore, this type of error can materialize irrespective of the computer or computational environment.
Triangulation is a prerequisite for FEMs, and its correctness is essential in obtaining valid finite element solutions.
There is no guarantee that a finite element solution obtained on an incorrect triangulation provides a reasonable approximation of the exact solution to the target problem.
The primary objective of this paper is to present a verification algorithm that reliably alerts users to these triangulation errors.

\begin{figure}[H]
 \centering
 \includegraphics[keepaspectratio, scale=0.35]{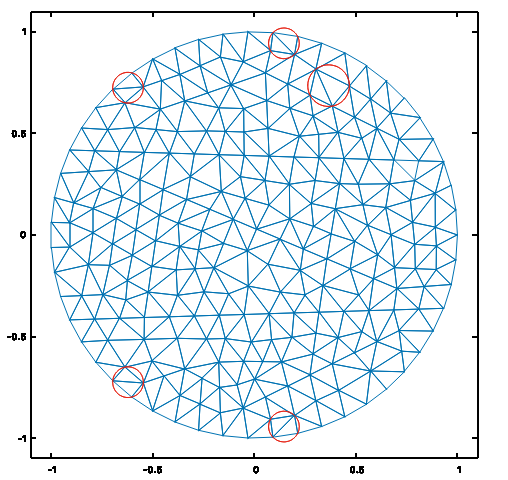}
 \caption{Triangulation obtained from the `buildmesh' function in FreeFEM++ v4.9. The areas highlighted by the circles do not satisfy the local Delaunay property.}
 \label{fig:wrong-delaunay}
\end{figure}
This paper focuses on guaranteeing the correctness and Delaunay property of triangulation. Correct Delaunay triangulation is vitally important for FEM. The minimum interior angle maximality (commonly referred to as the Delaunay property) of triangulation not only assists in obtaining numerical solutions with fewer errors, but is also important in preserving several properties of the original problem before discretization. For instance, the maximum principle applies to finite element solutions as well as solutions to the original continuous problem \cite{fujii1973some,knabner2003numerical,strang2008analysis}. 
In this context, it is crucial to ensure both the correctness of a triangulation and its Delaunay property. However, existing software that purports to return the Delaunay triangulation might output a triangulation that lacks the Delaunay property (see, for instance, Fig.~\ref{fig:wrong-delaunay}). Another aim of our study is to rectify such a triangulation so that it exhibits the Delaunay property.
An outstanding contribution of our research is its applicability to computer-assisted proofs based on finite element methods. 
Computer-assisted proofs aim to find an approximate solution to a target problem and establish the existence of an exact solution in the vicinity of this approximate solution with an explicit error bound (see, for instance, \cite{nakaoplumwatanabe2019numerical} and the references therein). 
The correctness of the triangulation is an imperative prerequisite for ensuring the accuracy of such a ``proof.'' 
Moreover, determining the smallest possible interpolation error constant $C_h$ (where $h$ symbolizes the mesh size) is critical in computer-assisted proofs. 
For the Dirichlet problem of an elliptic equation, we require an explicit value of $C_h$ that satisfies $\left\|v-P_h v\right\|_{L^2} \leq C_h\left\|v-P_h v\right\|_{V} $ for all $ v \in V$, where $P_h$ denotes the orthogonal projection of the proper solution space $V$ to a finite element space $V_h$. 
For additional details, see \cite{nakaoplumwatanabe2019numerical}. 
The Delaunay property of the triangulation is beneficial because it minimizes $C_h$ for a fixed set of vertices. 
Furthermore, the Delaunay property enables us to determine how the vertices should be positioned to ensure that $C_h$ remains below a desired value (see, for example, \cite{kobayashi2015circumradius}).
The objectives of this paper can be summarized as follows. First, we validate the correctness of a triangulation output using a given software package {\emph post hoc}. Second, we verify the Delaunay property of the correct triangulation and make corrections if this property is not satisfied. A general method for rectifying errors in the triangulation itself, such as overlapping triangles, is extremely challenging, and is not covered in this paper. However, it is possible to identify the location of errors. By removing or slightly moving the points in these areas, seriously erroneous results produced by rounding errors can be avoided, increasing the chances of obtaining a correct triangulation.
Such manipulations hold potential benefits, especially when applied in the context of FEM, where the precision and accuracy of triangulation are crucial for reliable analysis and simulations.
Although there are robust libraries for 2D Delaunay triangulation, such as Shewchuk's Triangle and CGAL, there is no guarantee that all existing software always generate correct Delaunay triangulations, and erroneous results may be output under the influence of rounding errors and human mistakes.
Even if a particular software package could produce perfectly accurate triangulations, it would be unrealistic to expect all existing software to operate in this manner. It is useful to have the flexibility of employing any mesh-generating algorithm, especially one that is already familiar. There is also the matter of the vast triangulation data that have already been generated. Being able to provide {\emph a posteriori} guarantees is crucial for these reasons.
To ensure the correctness of triangulation, we propose the Polygonal Sequence-driven Triangulation Validator (PSTV).
One straightforward approach for verifying whether the entire target area is covered by triangles without any overlaps is to examine the absence of overlaps for every possible pair of triangles in the dataset.
For a dataset with $n$ triangles, this would entail $n(n-1)/2$ comparisons. In computational terms, this results in a time complexity that is proportional to the square of the number of triangles. Hence, conducting such a verification would rapidly become computationally intensive as the number of triangles increases.
The fundamental design of PSTV incrementally generates a sequence of polygons whose interior forms a triangulation. This significantly reduces the computational complexity and time of verification, achieving an average computational order of $n^{1.6}$ for a given number of nodes $n$ according to numerical experiments (see Section~\ref{sec:experiments}).
Moreover, a distinctive feature of PSTV is that it does not employ interval arithmetic, thus avoiding the effects of rounding errors during execution. Interval arithmetic \cite{moore1966interval,sunaga1958theory} is a simple and effective method for evaluating rounding errors, but if all computations are replaced with interval arithmetic, the necessary computation time increases dramatically. To circumvent this issue, we use floating-point filters, which provide a sufficient condition for the correctness of the sign of the computed result at potentially low cost.
PSTV uses the filter proposed by Ozaki et al.~\cite{ozaki-filter} for the orientation test, which determines whether two line segments intersect. Moreover, we propose a new filter that functions reliably even when an underflow occurs during the incircle test, which determines whether a point lies inside the circumcircle of a triangle. 
This is an extension of Shewchuk's filter~\cite{shewchuk-filter}.
If the floating-point filters cannot verify the correctness of the sign, we apply a robust algorithm (in our implementation example, rational number computation) to compute the sign of the determinant. The failure of floating-point filters to correctly determine the sign is extremely rare (occurring in less than 0.1\% of the computations in many examples). Hence, the use of the robust algorithm has a minimal effect on the total computation time.
The remainder of this paper is organized as follows. In Section \ref{sec:prep}, we present several definitions and symbols used throughout the paper, along with the three fundamental tests required to ensure triangulation. Additionally, we introduce the floating-point filters required to accurately perform these tests. In Section \ref{sec:pstv}, we introduce the PSTV algorithm, which is the primary subject of this paper. In Section \ref{sec:veridp}, we elaborate on the verification and corrections of the Delaunay property of triangulation ensured by PSTV. Section \ref{sec:experiments} applies the PSTV method and the algorithm for verifying and correcting the Delaunay property to several specific domains, enabling an evaluation of the proposed methods. Finally, we summarize the results of this study in Section \ref{sec:conc}.
\section{Preparation}\label{sec:prep}
This study focuses on cases in which the coordinates of nodes in the triangulation are expressed as binary floating-point numbers, referred to here as the set $\mathbb{F}$.
Our main goal is to verify the validity of certain datasets as accurate representations of the triangulation under consideration.
The datasets in question are as follows:
\begin{enumerate}
\item $\mathbb{S}=\{p_i\in \mathbb{F}^2 \mid i=1,\ldots,n_p\}
$: This set consists of unique input nodes such that $p_i \neq p_j (i \neq j)$ is satisfied, each comprising at least two elements from the binary floating-point numbers.
\item $\mathbb{T}=\{T_i\in \mathbb{S}^3 \mid i=1,\ldots,n_t\}$: A set of triangles, with each triangle $T_i$ formed using three nodes from $\mathbb{S}$.
\item $\mathbb{B}$: A sequence of boundary nodes, ordered in a clockwise direction.
\end{enumerate}
These datasets are maintained on a computer as follows: $\mathbb{S}$ is stored as a $2\times n_p$ array of binary floating-point numbers, $\mathbb{T}$ is stored as a $3\times n_t$ array of node indices, and $\mathbb{B}$ is stored as a 1D array of integers of a certain length.
We represent the oriented line segment from point $p$ to $q$ in $\mathbb{R}^2$ as $\overrightarrow{pq}$, and the circumcircle of a triangle formed by three points $p$, $q$, and $r$ in counterclockwise order as $C(p,q,r)$.
In the following sections, we introduce three fundamental tests involving floating-point operations. These tests are necessary for verification purposes.
\subsection{Rigorous Computation for Precise Tests}\label{subsec:rt}
In calculations involving the orientation and incircle tests, there is the risk of rounding errors occurring during finite precision floating-point arithmetic. To prevent such inaccuracies, it is essential to employ rigorous computation. 
However, these tests are frequently performed when verifying the correctness of a triangulation and its Delaunay property. Using interval arithmetic for all calculations can significantly increase the computation time. 
As such, we use floating-point filters to determine the precision of the calculation results based on an approximation calculation using floating-point arithmetic. We resort to rational arithmetic with the GNU Multiple Precision (GMP) library when the accuracy of the tests cannot be assured by floating-point filters alone. Although the use of rational arithmetic with GMP entails a substantial computational cost, experiments have shown that instances where the correctness of the tests cannot be guaranteed by floating-point filters constitute less than 0.1\% of the total; thus, they do not significantly impact the overall computation time (see Section \ref{sec:experiments}).
Several floating-point filters have been proposed for checking the sign of the determinant of a matrix (e.g., \cite{shewchuk-filter,burnikel2001exact,devillers2003efficient,sharma2017robust,ozaki2009adaptive,graillat2005applications,demmel2004fast}).
In this study, Ozaki's floating-point filter \cite{ozaki-filter} is used for orientation tests, whereas Shewchuk's floating-point filter~\cite{shewchuk-filter} is used for incircle tests.
The original form of Shewchuk's filter~\cite{shewchuk-filter} does not consider underflow, so we use an extended filter that takes underflow into account, thereby ensuring 100\% accurate test results.
The proposed PSTV algorithm assumes that all nodes of the dataset constituting the triangulation are represented as floating-point numbers as defined in IEEE 754-2008~\cite{ieee}.
Let $\textbf{u}$ denote the rounding unit, which is, for instance, $2^{-53}$ for binary64.
Let $\textbf{u}_N$ represent the smallest positive normalized floating-point number, for instance, $2^{-1022}$ for binary64.
When using Ozaki's floating-point filter \cite[Algorithm 3]{ozaki-filter}, a rigorous orientation test can be achieved using Algorithm \ref{alg:verified-ot}.
Using the filter based on Shewchuk's floating-point filter, a rigorous incircle test can be performed using Algorithm \ref{alg:verified-ict}.
\subsection{Orientation test}
The orientation test $(\OT)$ determines the position relationship between a point and an oriented line segment.
Given three points $p_a(x_a,y_a)$, $p_b(x_b,y_b)$, and $p_c(x_c,y_c)$ in $\mathbb{R}^2$, the orientation test determines whether $p_c$ is to the left or right of the oriented line segment $\overrightarrow{p_ap_b}$, or if it lies on the line.
To do this, we define the function $\OT (p_a,p_b,p_c)$ as follows:
\[
\OT (p_a,p_b,p_c) = 
\begin{vmatrix}
x_a-x_c & y_a-y_c\\
x_b-x_c & y_b-y_c\\
\end{vmatrix}
\]
Then, the following conditions hold:
\[
\OT (p_a,p_b,p_c)
\begin{cases}
    > 0 \Rightarrow \text{$p_c$ is to the left of $\overrightarrow{p_ap_b}$}\\
    < 0 \Rightarrow \text{$p_c$ is to the right of $\overrightarrow{p_ap_b}$}\\
    = 0 \Rightarrow \text{$p_c$ is on $\overrightarrow{p_ap_b}$}\\
\end{cases}
\]
The calculation of $\OT$ is subject to the influence of rounding errors, and these must be taken into account. Detailed measures for handling this issue are elaborated in Subsection~\ref{subsec:rt}.
\begin{figure}[ht]
    \centering
    \includegraphics[keepaspectratio, scale=0.4]{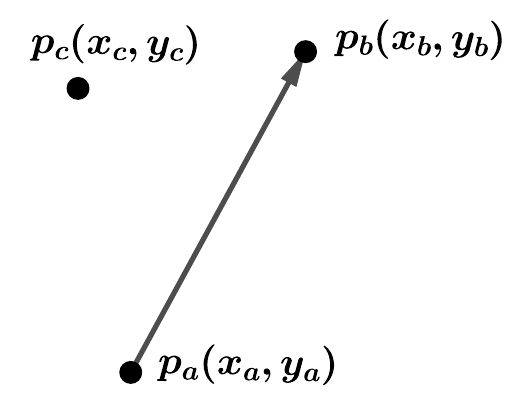}
    \caption{Example of orientation test (point $p_c$ is to the left of the oriented line segment $\overrightarrow{p_a p_b}$).}
    \label{fig:orientation}
\end{figure}
\begin{algorithm}[H]
\caption{OT: Orientation Test}
\label{alg:verified-ot}
\begin{algorithmic}
\Function{OT}{Points($p_a(x_a, y_a),~p_b(x_b, y_b),~p_c(x_c, y_c)$)}
\State $l \gets (x_a - x_c)*(y_b - y_c)$
\State $r \gets (x_b - x_c)*(y_a - y_c)$
\State $det \gets l - r$
\State $criteria \gets \theta * (|l+r|+ \textbf{u}_N)$ //$\theta=3u$
\If {$|det|>criteria$}
\State \Return $det$
\Else
\State rationally calculate $det$ with GMP
\State \Return sgn($det$)
\EndIf
\EndFunction
\end{algorithmic}
\end{algorithm}
\subsection{Incircle test}
The incircle test  $(\ICT)$ determines the position of a point relative to the circumcircle of a triangle.
Given four points $p_a(x_a,y_a)$, $p_b(x_b,y_b)$, $p_c(x_c,y_c)$, $p_d(x_d,y_d)$ in $\mathbb{R}^2$, where $p_a,p_b,p_c$ are in counterclockwise order, the incircle test determines whether $p_d$ lies inside or outside the circumcircle $C(p_a,p_b,p_c)$, or if it lies on the circumcircle. 
It is assumed that $p_a$, $p_b$, and $p_c$ are not collinear.
We define the function $\ICT (p_a,p_b,p_c,p_d)$ as follows:
\[
\ICT (p_a,p_b,p_c,p_d) = 
\begin{vmatrix}
1 & x_a & y_a & x_a^2+y_a^2\\
1 & x_b & y_b & x_b^2+y_b^2\\
1 & x_c & y_c & x_c^2+y_c^2\\
1 & x_d & y_d & x_d^2+y_d^2\\
\end{vmatrix}
\]
Then, the following conditions hold:
\[
\ICT (p_a,p_b,p_c)
\begin{cases}
    > 0 \Rightarrow \text{$p_d$ is outside the circle $C(p_a,p_b,p_c)$}\\
    < 0 \Rightarrow \text{$p_d$ is inside the circle $C(p_a,p_b,p_c)$}\\
    = 0 \Rightarrow \text{$p_d$ is on the circle $C(p_a,p_b,p_c)$}\\
\end{cases}
\]
The outcome of this test is also susceptible to the influence of rounding errors. Detailed strategies for handling these issues are discussed in Subsection~\ref{subsec:rt}.
\begin{figure}[ht]
    \centering
    \includegraphics[keepaspectratio, scale=0.4]{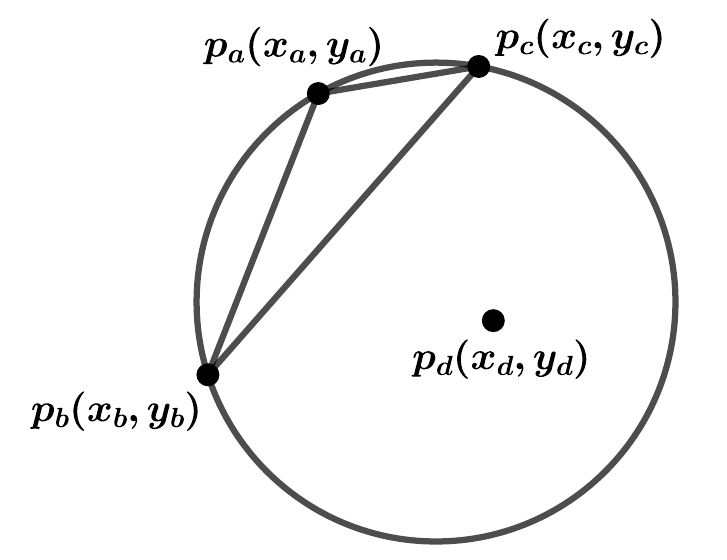}
    \caption{Example of incircle test (point $p_d$ lies inside the circumcircle of triangle $p_a p_b p_c$).}
    \label{fig:incircle}
\end{figure}
\begin{algorithm}[H]
\caption{ICT: Incircle Test}
\label{alg:verified-ict}
\begin{algorithmic}
\renewcommand{\algorithmicrequire}{\textbf{Input:}}
\Require 
Points($p_a(x_a, y_a),~p_b(x_b, y_b),~p_c(x_c, y_c),~p_d(x_d, y_d)$)
\State $adx \leftarrow x_a - x_d$, \quad $bdx \leftarrow x_b - x_d$, \quad $cdx \leftarrow x_c - x_d$
\State $ady \leftarrow y_a - y_d$, \quad $bdy \leftarrow y_b - y_d$, \quad $cdy \leftarrow y_c - y_d$
\State $\alpha_{a1} \leftarrow adx^2+ady^2$, \quad $\alpha_{a2} \leftarrow bdx*cdy-bdy*cdx$, \quad $\alpha_{a2'} \leftarrow |bdx*cdy|+|bdy*cdx|$
\State $\alpha_{a} \leftarrow \alpha_{a1}*\alpha_{a2}$
\State $\alpha_{a'} \leftarrow \alpha_{a1}*\alpha_{a2'}$
\State $\alpha_{b1} \leftarrow bdx^2+bdy^2$, \quad $\alpha_{b2} \leftarrow cdx*ady-cdy*adx$, \quad $\alpha_{b2'} \leftarrow |cdx*ady|+|cdy*adx|$
\State $\alpha_{b} \leftarrow \alpha_{b1}*\alpha_{b2}$
\State $\alpha_{b'} \leftarrow \alpha_{b1}*\alpha_{b2'}$
\State $\alpha_{c1} \leftarrow cdx^2+cdy^2$, \quad $\alpha_{c2} \leftarrow adx*bdy-ady*bdx$, \quad $\alpha_{c2'} \leftarrow |adx*bdy|+|ady*bdx|$
\State $\alpha_{c} \leftarrow \alpha_{c1}*\alpha_{c2}$
\State $\alpha_{c'} \leftarrow \alpha_{c1}*\alpha_{c2'}$
\State $det \leftarrow \alpha_{a}+\alpha_{b}+\alpha_{c}$
\State $\beta_{a} \leftarrow \alpha_{a1}*\alpha_{a2'}$
\State $\beta_{b} \leftarrow \alpha_{b1}*\alpha_{b2'}$
\State $\beta_{c} \leftarrow \alpha_{c1}*\alpha_{c2'}$
\State $errbound \leftarrow (10*u+176*u^2)*(\beta_{a}+\beta_{b}+\beta{c}) + 3*u_s*((\alpha_{a2'} + \alpha_{a1}) + (\alpha_{b2'}+ \alpha_{b1} ) + (\alpha_{c2'} + \alpha_{c1} ) + 1)$
\If{$|det|>errbound$}
\State \Return $det$
\Else
\State rationally calculate $det$ with GMP
\State \Return $det$
\EndIf
\end{algorithmic}
\end{algorithm}
\subsection{Intersection test }
\begin{figure}[H]
    \centering
    \begin{tabular}{cc}
        \begin{minipage}[t]{0.38\hsize}
            \centering
            \includegraphics[keepaspectratio, scale=0.55]{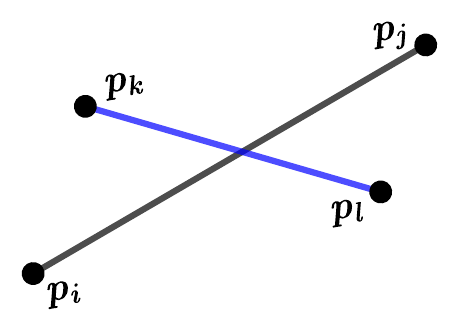}
            \subfloat{(a) intersect}
            \label{fig:ist1}
        \end{minipage}
        \begin{minipage}[t]{0.38\hsize}
            \centering
            \includegraphics[keepaspectratio, scale=0.55]{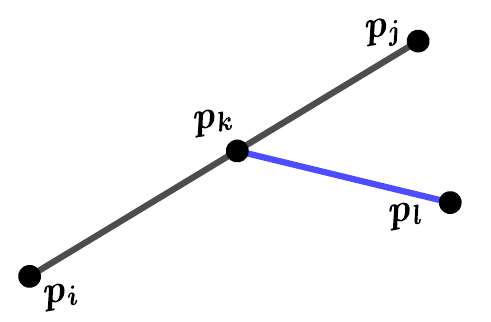}
            \subfloat{(b) intersect}
            \label{fig:ist2}
        \end{minipage}\\\\
        
        \begin{minipage}[t]{0.4\hsize}
            \centering
            \includegraphics[keepaspectratio, scale=0.45]{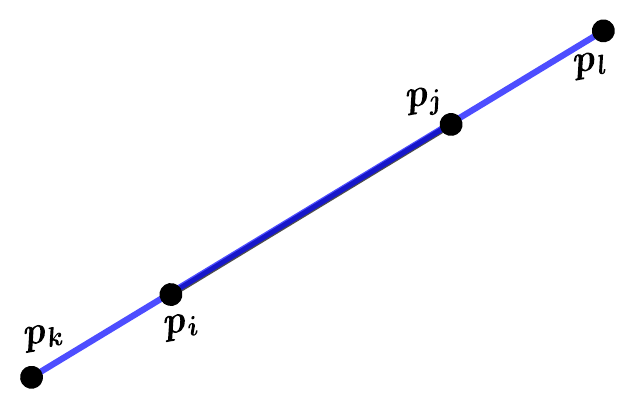}
            \subfloat{(c) intersect}
            \label{fig:ist3}
        \end{minipage}
        \begin{minipage}[t]{0.4\hsize}
            \centering
            \includegraphics[keepaspectratio, scale=0.45]{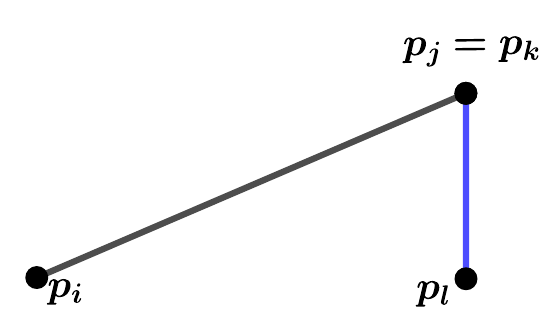}
            \subfloat{(d) not intersect}
            \label{fig:ist4}
        \end{minipage}
    \end{tabular}
    \caption{Geometric pattern for intersection test of two line segments.}
    \label{fig:ist-pattern}
\end{figure}

The intersection test $(\IST)$ determines whether two line segments intersect.
The criterion for determining whether two line segments intersect is defined by the pattern shown in Fig.~\ref{fig:ist-pattern}.
This process constitutes a computationally intensive task within the entire PSTV algorithm, because the process is iteratively invoked during Step 2.
Given four points $p_a(x_a,y_a)$, $p_b(x_b,y_b)$, $p_c(x_c,y_c)$, and $p_d(x_d,y_d)$ in $\mathbb{R}^2$, the intersection test ascertains whether the line segments $\overline{p_a p_b}$ and $\overline{p_c p_d}$ intersect. We define the line segments as non-intersecting if one endpoint of a line segment matches an endpoint of the other line segment.
First, we introduce the onsegment test (\OST), which checks whether a point $p_c$ lies on a given line segment $\overline{p_a p_b}$. 
\begin{algorithm}[ht]
\caption{\OST: Onsegment Test}
\label{alg:ost}
\begin{algorithmic}[1]
\renewcommand{\algorithmicrequire}{\textbf{Input:}}
\Function{\OST}{Segment($p_a, p_b$) and Point($p_c$)}
\If {$\OT(p_a, p_b, p_c)==0$}
\If {${\rm Min}(x_a,x_b)<x_c ~\&~ x_c<{\rm Max}(x_a,x_b) ~\&~ {\rm Min}(y_a,y_b)<y_c ~\&~ y_c<{\rm Max}(y_a,y_b)$}
\State \Return TRUE
\EndIf
\EndIf
\State \Return FALSE
\EndFunction
\end{algorithmic}
\end{algorithm}

\newpage
The onsegment test is used to execute the intersection test. Initially, we designate the line passing through the endpoints of one line segment as the boundary line and evaluate whether the other endpoint of the line segment lies on both sides of this boundary line. If it does, we conclude that the two line segments intersect. Four precise $\OT$s can ensure accurate intersection detection. If an endpoint resides on the boundary line, we employ the onsegment test to ascertain whether the endpoint lies on the other line segment excluding its endpoints. If it does, we determine that the two line segments intersect.
\begin{algorithm}[H]
\caption{\IST: Intersection Test}
\label{alg:intersection}
\begin{algorithmic}[1]
\renewcommand{\algorithmicrequire}{\textbf{Input:}}
\Function{\IST}{Points($p_a, p_b, p_c, p_d$)}
\State ${\rm O}_{abc} \gets \OT (p_a, p_b, p_c)$
\State ${\rm O}_{abd} \gets \OT (p_a, p_b, p_d)$
\State ${\rm O}_{cda} \gets \OT (p_c, p_d, p_a)$
\State ${\rm O}_{cdb} \gets \OT (p_c, p_d, p_b)$
\State \Comment{In the actual algorithm, there are cases where the result can be determined with only two $\OT$s instead of four $\OT$s.}
\If {${\rm O}_{abc}*{\rm O}_{abd}<0 ~\&~ {\rm O}_{cda}*{\rm O}_{cdb}<0$}
\State \Return TRUE //Intersect
\EndIf
\If {$\OST(\it p_a, p_b, p_c)$}
\State \Return TRUE //Intersect
\EndIf
\If {$\OST(\it p_a, p_b, p_d)$}
\State \Return TRUE //Intersect
\EndIf
\If {$\OST(\it p_c, p_d, p_a)$}
\State \Return TRUE //Intersect
\EndIf
\If {$\OST(\it p_c, p_d, p_b)$}
\State \Return TRUE //Intersect
\EndIf
\State \Return FALSE
\EndFunction
\end{algorithmic}
\end{algorithm}

\section{Polygonal Sequence-driven Triangulation Validator}\label{sec:pstv}
\begin{figure}[ht]
    \centering
    \includegraphics[keepaspectratio, scale=0.6]{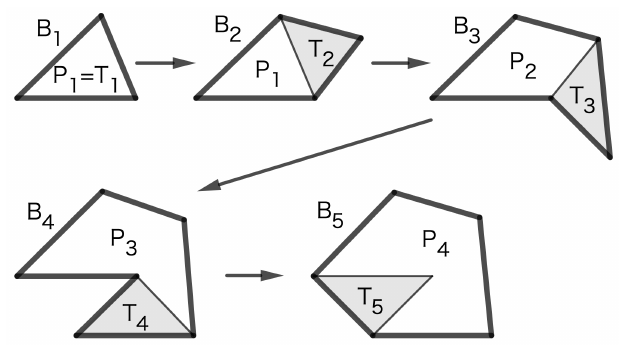}
    \caption{Illustration of the incremental construction of polygonal and boundary sequences for triangulation verification.}
    \label{fig:triangulation_process}
\end{figure}

The PSTV algorithm is an innovative method for efficiently validating the correctness of a triangulation dataset. The cornerstone of its design is the incremental generation of a sequence of triangulations, thereby considerably reducing the computational complexity and time of the verification process.
The triangulation verification process entails confirming that the triangles in a dataset cover the target region without overlap. However, in the absence of appropriate optimization, the computation time required for this verification increases in proportion to the square of the number of triangles. Moreover, the implementation of interval operations could potentially inflate the computational cost. In light of these challenges, the PSTV algorithm emerges as an optimized solution process.
As described in Section \ref{sec:prep}, the input to the PSTV algorithm includes the set of unique nodes $\mathbb{S}$ composed of three or more points (represented in binary), the set of triangles $\mathbb{T}$ formed by selecting three points from $\mathbb{S}$, and the boundary node sequence $\mathbb{B}$ obtained by arranging the nodes of $\mathbb{S}$ to form the outer perimeter.
The PSTV algorithm first selects an initial triangle $T_1$ from the input set to serve as the initial polygon $P_1$ for the polygonal sequence. When selecting $T_1$, the determination of whether $T_1$ is a triangle is carried out using an orientation test. The algorithm then progressively constructs a sequence of polygons $P_k$ and their corresponding boundary sequences $B_k$ by iteratively connecting adjacent triangles, as depicted in Fig.~\ref{fig:triangulation_process}.
The process of constructing the polygonal sequence and boundary sequence comprises the following steps:\\

\begin{enumerate}
\item[\textbf{Step 1.}] Identify an adjacent triangle $T_{k+1}$ of $P_k$.
\item[\textbf{Step 2.}] Evaluate whether the dataset formed by $P_k \cup T_{k+1}$ constitutes a correct triangulation.
\item[\textbf{Step 3.}] If the dataset is valid, we generate $P_{k+1}$ as $P_k \cup T_{k+1}$ and denote the boundary set of $P_{k+1}$ as $B_{k+1}$. If not, terminate the process.\\
\end{enumerate}
Following the completion of Steps 1--3, the process returns to Step 1 and the cycle continues. By iterating this process until all input triangles have been incorporated into the polygonal sequence, and as long as the final polygonal boundary $B_k$ aligns with the input boundary $\mathbb{B}$, the PSTV effectively confirms the correct triangulation of the input dataset.
In this context, the equivalence of $B_k$ to $\mathbb{B}$ implies that the number of nodes in each respective sequence is the same. Furthermore, when considering $B_k = (a_i)$ and $\mathbb{B} = (b_i)$, where $i = 1, \ldots, n_b$, the condition is satisfied if
\[
\exists k\, \text{s.t.} \forall i, a_i = b_{i+k\,mod\,n_b}
\]
Here, $m_1\mod m_2 = r$ signifies that $r$ is the remainder when $m_1$ is divided by $m_2$.
This transformative approach allows for the efficient and optimized verification of a triangulation, resulting in significant reductions in both computation time and complexity.
In the following, we analyze each of these steps and elucidate the nuances and considerations of the PSTV algorithm. 

\subsubsection*{Step 1: Efficient Search of Adjacent Triangles}
\begin{figure}[H]
    \centering
    \includegraphics[keepaspectratio, scale=0.4]{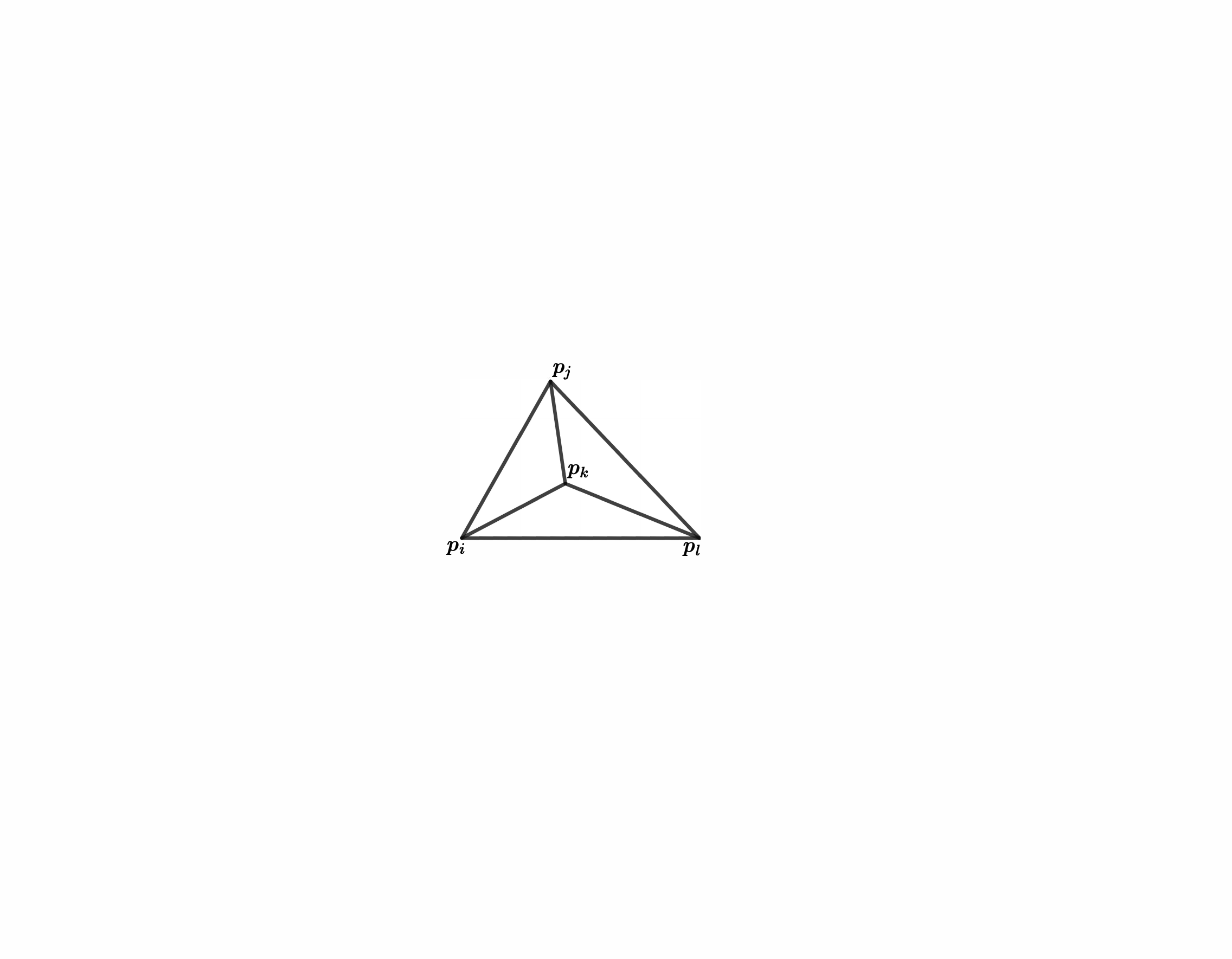}
    \caption{Example of triangulation.}
    \label{fig:edgemap}
\end{figure}

When searching for a triangle adjacent to the polygon ${P_k}$, we first select one boundary edge from the set of boundary edges of ${P_k}$. The aim is to identify triangles that include this selected edge. As the number of triangles in the input dataset increases, searching through all triangles in every iteration could cause the computation time to escalate rapidly.
To overcome this computational challenge, we propose the creation of an edge map. This edge map is a data structure that records all triangles associated with each edge, essentially serving as an index for faster search and retrieval. The edge map thus provides rapid access to the set of triangles adjacent to a given boundary edge. 
In theory, the number of triangles associated with a given edge is always one or two. If the edge forms a boundary of the target region, then there will be only one such triangle. For all other edges, there will be two triangles. Constructing the edge map has a computational complexity of $O(n_t)$. Once constructed, the edge map enables triangles adjacent to the edge to be searched with a computational complexity of $O(1)$.
To illustrate, consider the dataset shown in Fig.~\ref{fig:edgemap}. The only triangle that includes edge $\overline{p_i p_j}$ is $\triangle{p_i p_j p_k}$, while there are two triangles, $\triangle{p_i p_j p_k}$ and $\triangle{p_k p_l p_i}$, which include edge $\overline{p_i p_k}$. Considering all edges, the resulting edge map would be:
\[
\begin{aligned}
\begin{Bmatrix}
\overline{p_i p_j} & : & \{\triangle{p_i p_j p_k}\}\\
\overline{p_i p_k} & : & \{\triangle{p_i p_j p_k},\triangle{p_k p_l p_i}\}\\
\overline{p_i p_l} & : & \{\triangle{p_k p_l p_i}\}\\
\overline{p_j p_k} & : & \{\triangle{p_i p_j p_k},\triangle{p_j p_k p_l}\}\\
\overline{p_j p_l} & : & \{\triangle{p_j p_k p_l}\}\\
\overline{p_k p_l} & : & \{\triangle{p_j p_k p_l},\triangle{p_k p_l p_i}\}\\
\end{Bmatrix}
\end{aligned}
\]
If the edge map shows that only one triangle corresponds to a given boundary edge, that triangle is considered the adjacent triangle. If there are two corresponding triangles, one of them must already be included in $P_k$, so the other triangle is deemed the adjacent triangle.
By employing this edge map as a preprocessing step for the triangulation verification algorithm, we can rapidly locate an adjacent triangle to a boundary edge in polygon $P_k$.

\subsubsection*{Step 2: Evaluating the Correctness of the Triangulation}
\begin{figure}[H]
    \centering
    \begin{tabular}{cc}
        \begin{minipage}[t]{0.38\hsize}
            \centering
            \includegraphics[keepaspectratio, scale=0.4]{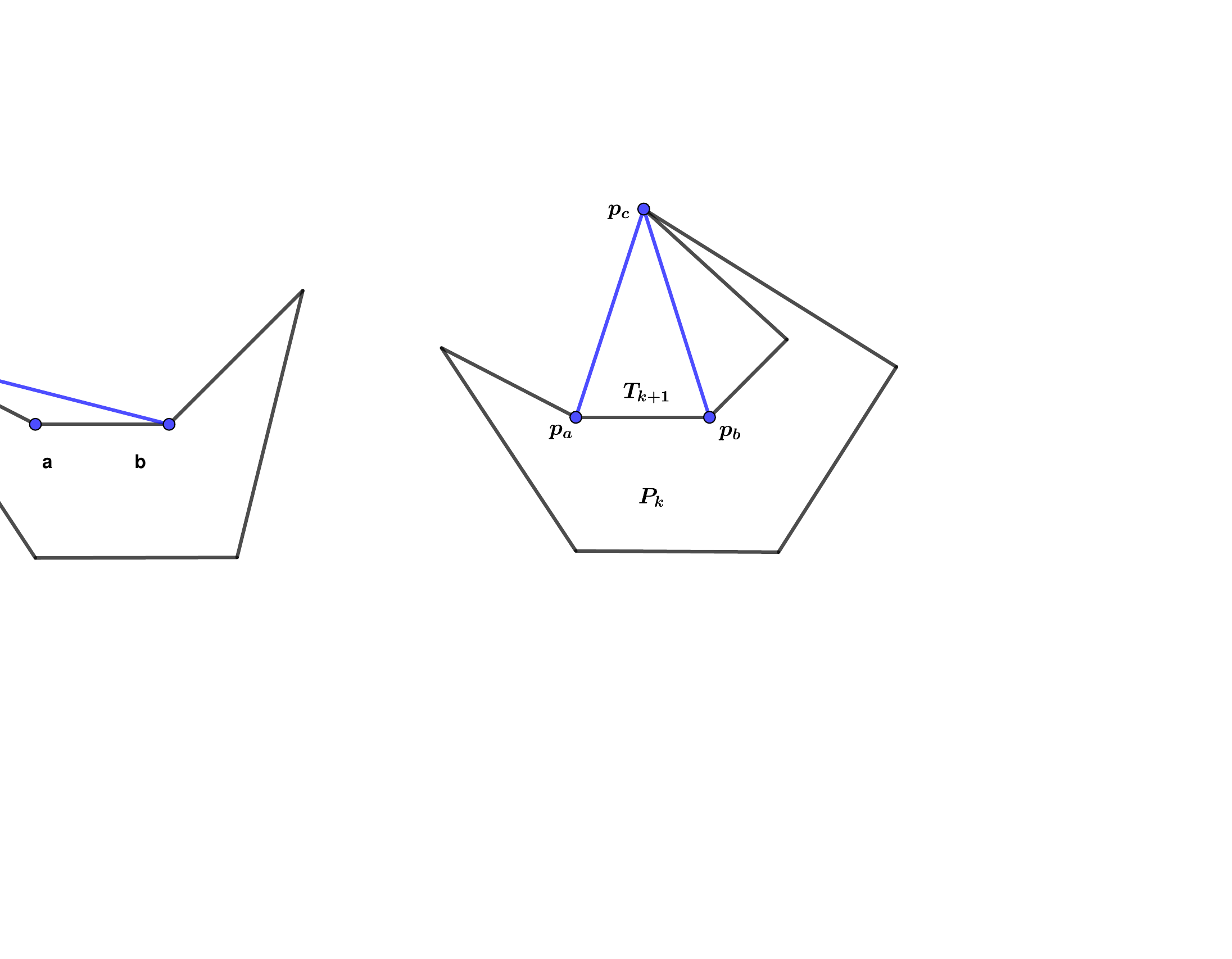}
            \subfloat{($\mathrm{i}$) 3 points 1 edge shared}
            \label{fig:pattern3p1e}
        \end{minipage}
        \begin{minipage}[t]{0.38\hsize}
            \centering
            \includegraphics[keepaspectratio, scale=0.4]{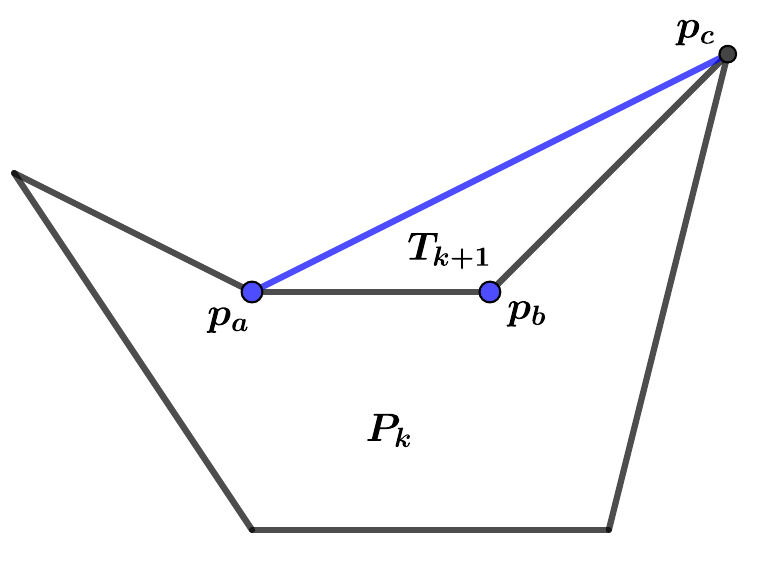}
            \subfloat{($\mathrm{ii}$) 3 points 2 edge shared}
            \label{fig:pattern3p2e}
        \end{minipage}\\\\
        
        \begin{minipage}[t]{0.4\hsize}
            \centering
            \includegraphics[keepaspectratio, scale=0.4]{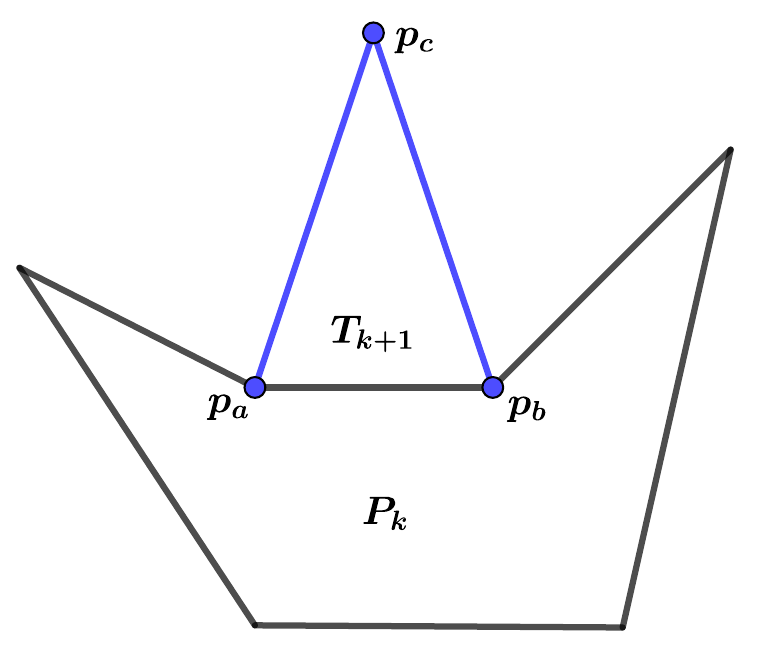}
            \subfloat{$(\mathrm{iii}$) 2 points 1 edge shared}
            \label{fig:pattern2p1e}
        \end{minipage}
        \begin{minipage}[t]{0.4\hsize}
            \centering
            \includegraphics[keepaspectratio, scale=0.4]{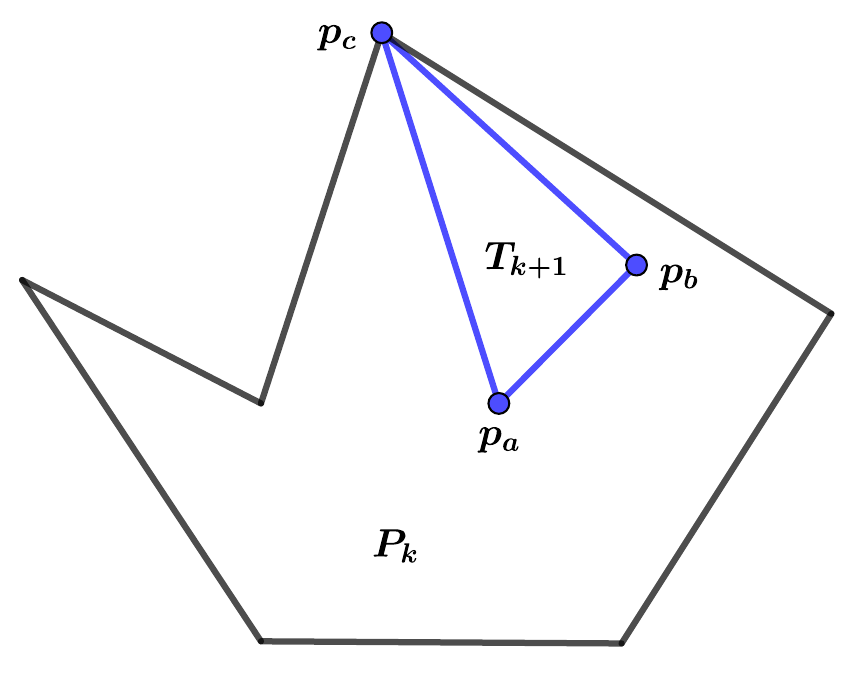}
            \subfloat{$(\mathrm{iv})$ 3 points 3 edge shared}
            \label{fig:pattern3p3e}
        \end{minipage}
    \end{tabular}
    \caption{Adjacency patterns of polygons and triangles that can occur in the case of correct triangulation.}
    \label{fig:adjacent-pattern}
\end{figure}

There exist various configurations for the adjacency of a triangle, $T_{k+1} = \Delta p_a p_b p_c$, with respect to a boundary edge $\overline{p_a p_b}$ of a polygon $P_k$. These adjacency configurations can be grouped into four distinct categories, based on the number of shared points (up to 3) and shared edges (up to 3). An instance of such adjacency, where a polygon and a triangle share 3 points and 1 edge, is depicted in Fig.~\ref{fig:adjacent-pattern}(i). This particular state is referred to as ``3 points 1 edge shared'' adjacency. When the triangulation is correct, the adjacency methods are limited to the four patterns of Fig.~\ref{fig:adjacent-pattern}. The four patterns are:
\begin{enumerate}
\item[($\mathrm{i}$)] 3 points 1 edge shared
\item[($\mathrm{ii}$)] 3 points 2 edges shared
\item[($\mathrm{iii}$)] 2 points 1 edge shared
\item[($\mathrm{iv}$)] 3 points 3 edges shared
\end{enumerate}
Table \ref{tab:adjustPattern} describes these adjacency patterns between polygon boundary edges and triangles.
\begin{table}[H]
\centering
\caption{Adjacency patterns between polygon boundary edges and triangles.}
\label{tab:adjustPattern}
\begin{tabular}{|c|c|c|c|c|} \hline
Shared Edges ~\textbackslash~ Shared Points & 0 & 1 & 2   & 3   \\ \hline
0                      & - & - & -   & -   \\
1                      & - & - & (iii) & (i) \\
2                      & - & - & -   & (ii) \\
3                      & - & - & -   & (iv) \\ \hline
\end{tabular}
\end{table}
Category ($\mathrm{iv}$) corresponds to the scenario in which the adjacent triangle creates a single triangle-shaped hole. However, to prevent complications within the algorithm, we purposefully avoid such configurations. This particular choice is discussed in the subsequent description.

\subsubsection*{Category $(\mathrm{i})$: 3 Points 1 Edge Shared}
When the adjacent triangle $T_{k+1}$ is connected to the polygon $P_k$, it results in a configuration encompassing an area that has yet to be validated. In such a scenario, we forego this configuration, returning to Step 1 to identify a different adjacent triangle. The regions yet to be validated invariably yield a triangle configuration with 3 shared points and 3 shared edges. Therefore, we purposely avoid connecting a triangle with a category ($\mathrm{iv}$) 3 points 3 edges shared adjacency pattern.

\subsubsection*{Category $(\mathrm{ii})$: 3 Points 2 Edges Shared or Category $(\mathrm{iii})$: 2 Points 1 Edge Shared}
\begin{figure}[H]
    \centering
    \includegraphics[keepaspectratio, scale=0.25]{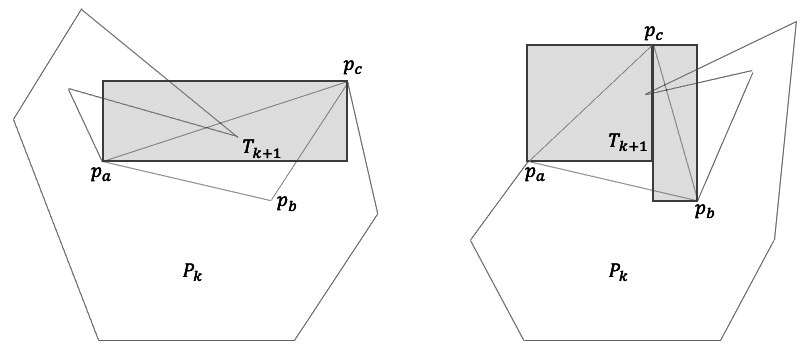}
    \caption{Illustration of the 3 points 2 edges shared and 2 points 1 edge shared adjacency patterns between polygon $P_k$ and adjacent triangle $T_{k+1}$.}
    \label{fig:adjust3221}
\end{figure}

\begin{figure}[H]
    \centering
    \includegraphics[keepaspectratio, scale=0.25]{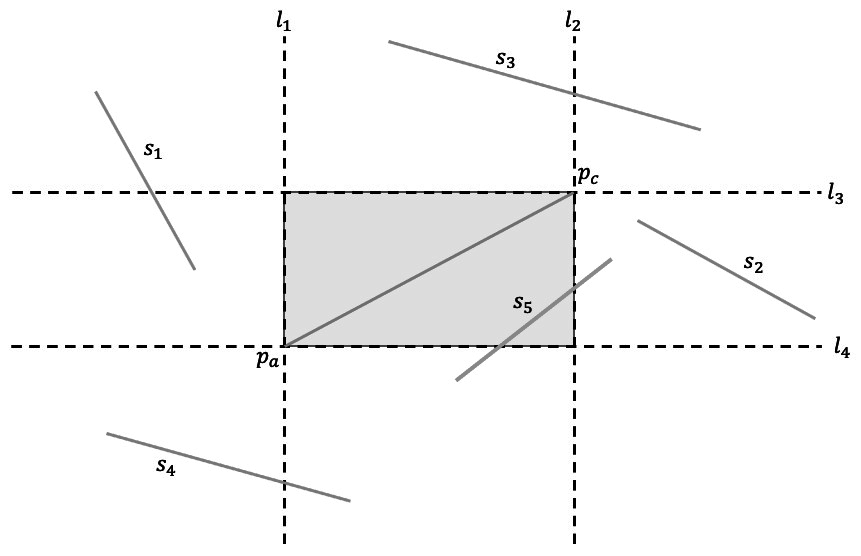}
    \caption{Illustration of edge intersection checks for optimizing boundary edge verification.}
    \label{fig:intersctionSegment}
\end{figure}

\begin{figure}[H]
    \centering
    \includegraphics[keepaspectratio, scale=0.35]{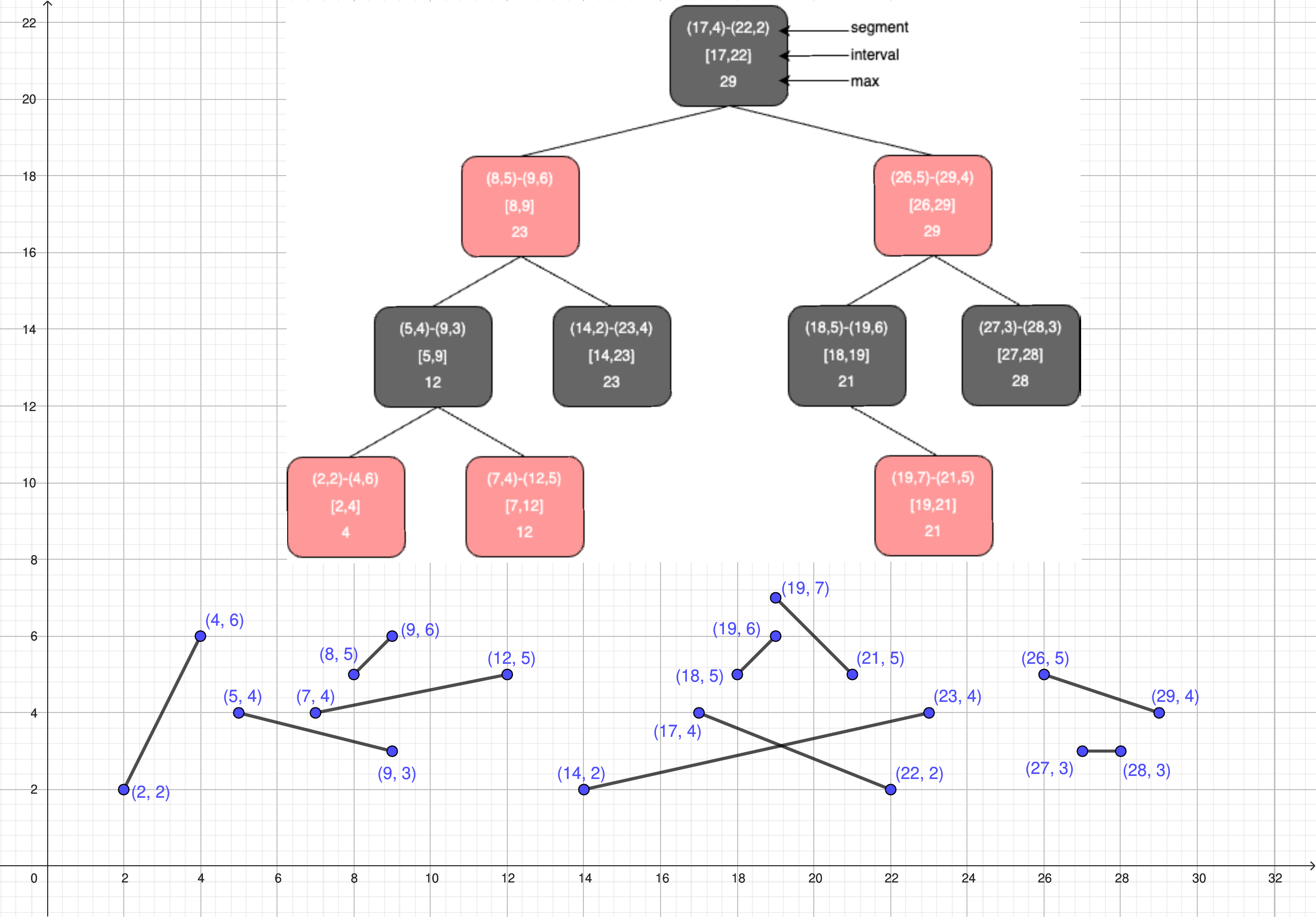}
    \caption{An interval tree, comprising the set of ten line segments located at the bottom of the image. In the case of an X-interval tree, each node contains information about the interval of the x-coordinates of endpoints, the maximum endpoint (described in the text), and the line segments. The entry in the root node represents the interval with inf-end point 17, sup-end point 22, maximum endpoint 29, and the line segment $(17,4)\mathrm{-}(22,2)$.}
    \label{fig:rb-tree}
\end{figure}

\begin{figure}[H]
    \centering
    \includegraphics[keepaspectratio, scale=0.4]{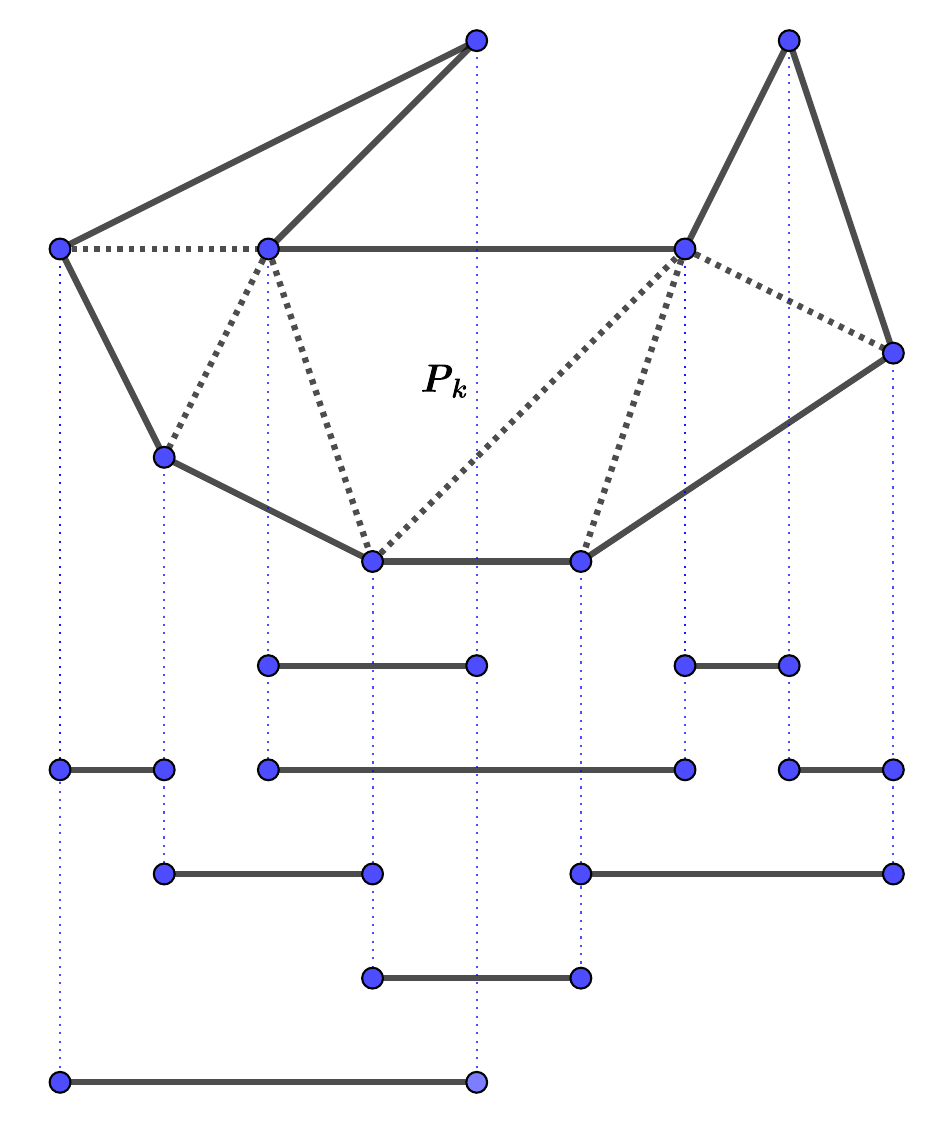}
    \caption{Intervals to be stored in the X-interval tree involve the projections onto the x-axis of the edges comprising the outer boundary of polygon $P_k$. Similarly, for the Y-interval tree, the intervals include the projections onto the y-axis of the edges forming the outer boundary of polygon $P_k$.}
    \label{fig:toXaxis}
\end{figure}

For the adjacency pattern of a triangle $T_{k+1}$ with 3 points 2 edges shared or 2 points 1 edge shared (as illustrated in Fig.~\ref{fig:adjust3221}), the following conditions must be checked:
Initially, we employ an orientation test [Algorithm~\ref{alg:verified-ot}] to ascertain whether point $p_c$ resides on the left of $\overrightarrow{p_a p_b}$. If $p_c$ lies on the right of $\overrightarrow{p_a p_b}$, the adjacent triangle would fall within the polygon $P_k$, resulting in overlap with other triangles and rendering the input dataset unsuitable for triangulation.
Subsequently, we ensure that the non-shared edge of the adjacent triangle does not intersect with any of the boundary edges of polygon $P_k$. However, conducting intersection tests for all boundary edges can be computationally demanding.
To optimize the intersection checks, we only apply the intersection test to the boundary edges present within the rectangle defined by using the non-shared edge as the diagonal. This methodology, along with an illustrative example, is provided in Fig.~\ref{fig:intersctionSegment}.
Algorithm \ref{alg:verifyAdjacencyTriangle} elucidates the procedure for verifying the adjacency of a triangle.
\begin{algorithm}[H]
\caption{Algorithm for Verifying Adjacent Triangle}
\label{alg:verifyAdjacencyTriangle}
\begin{algorithmic}[1]
\Function{VerifyAdjacentTriangle}{Adjacent triangle ($\Delta p_a p_b p_c$)}
\If {$OT(p_a,p_b,p_c)>0$}
\For {each unshared edge}
\State Compute $x_{min},y_{min},x_{max},y_{max}$
\For {each edge in boundary edges ($edge=\overline{p_i p_j}$)}
\If{Not intersecting with boundary}
\State \textbf{continue}
\EndIf
\If{IST(endpoints of unshared edge and boundary edge)}
\State \Return True //Triangulation has error.
\EndIf
\EndFor
\EndFor
\State \Return False
\Else
\State \Return True //Triangulation has error.
\EndIf
\EndFunction
\end{algorithmic}
\end{algorithm}
To efficiently determine the boundary edges on which the intersection test should be performed, we employ an interval tree. An interval tree is a tree-based data structure designed for storing intervals, enabling the efficient retrieval of all intervals that overlap with a specified query interval.
In this research, we use an extended interval tree based on a red--black tree. A red--black tree is a balanced binary search tree that allows search, insertion, and deletion operations to be performed in O($\log n$) time for a set of size $n$.
The order of nodes in this tree is determined based on the starting points (lower bounds) of each interval. 
Each node in the tree stores the interval and the maximum endpoint of all intervals in the entire subtree.
Let us consider the task of traversing all intervals that overlap with a given query interval in this interval tree. 
Here, an interval $[t_1, t_2]$ (where $t_1\leq t_2)$ represents the set $\{t \in \mathbb{R} | t_1 \leq t \leq t_2\}$. 
We denote a specific interval $[t_1, t_2]$ as $i$, with $i.\text{inf} = t_1$ and $i.\text{sup} = t_2$.
Two intervals $i$ and $i'$ are said to overlap if $i \cap i' \neq \emptyset$, which means that $i.\text{inf} \leq i'.\text{sup}$ and $i'.\text{inf} \leq i.\text{sup}$ must hold.
The intervals $i$ and $i'$ must always be in one of the following three states:
\begin{enumerate}
    \item $i$ and $i'$ overlap.
    \item $i.\text{sup} < i'.\text{inf}$.
    \item $i'.\text{sup} < i.\text{inf}$.
\end{enumerate}
Each node $x$ in interval tree $T$ stores an interval $x.interval$ and the maximum endpoint $x.max$ of all intervals stored in the subtree rooted at $x$. 
This is defined as follows, where $x.left$ and $x.right$ represent the left and right children of node $x$, respectively:
\[
x.max = \max(x.interval.sup, x.left.max, x.right.max)
\]
Additionally, information about the line segment is incorporated into these nodes (see Fig.~\ref{fig:rb-tree}).
Interval trees allow dynamic insertion and deletion \cite[Theorem 14.1]{cormen2022introduction}. In our method, the interval tree is used to ensure a rigorous intersection test. Through insertion and deletion operations, all line segments of the outer boundary of polygon $P_k$ are consistently stored in the X-interval tree ($T_x$) and Y-interval tree ($T_y$) (see Fig.~\ref{fig:toXaxis}).
Let $p(x_p, y_p)$ and $q(x_q, y_q)$ be points. The X-interval tree stores the interval $[x_p, x_q]$ composed of the x-coordinates of the endpoints when storing the line segment $\overline{pq}$ ($x_p \leq x_q$). Additionally, each node in the X-interval tree holds information about the line segment. The Y-interval tree has a similar structure.
To search for line segments that require intersection verification with the line segment $\overline{p_a p_c}$, the initial step is to use Algorithm \ref{alg:IntervalSearch} to explore all nodes in the X-interval tree that overlap with the interval $[x_{p_a}, x_{p_c}]$ and all nodes in the Y-interval tree that overlap with the interval $[y_{p_a}, y_{p_c}]$.
Finally, by searching for line segments that exist in both the nodes obtained from the X-interval tree's overlap search and those obtained from the Y-interval tree's overlap search, it is possible to identify line segments that require intersection verification with the line segment $\overline{p_a p_c}$ (see Algorithm \ref{alg:SegmentSearch}).
\begin{algorithm}[H]
\caption{Interval Search}
\label{alg:IntervalSearch}
\begin{algorithmic}[1]
\Function{Interval-Search}{$root, interval$}
\If{$root$ is \textbf{null}}
\State \Return
\EndIf
\If{$root.interval.inf\leq interval.sup$ \text{and} $root.interval.sup \geq interval.inf $}
\State \text{output}($root$) \Comment{Output overlapping intervals}
\EndIf
\If{$root.left \neq$ \textbf{null} and $root.left.max \geq interval.inf$}
\State \Call{Interval-Search}{$root.left, interval$}
\EndIf
\State \Call{Interval-Search}{$root.right, interval$}
\EndFunction
\end{algorithmic}
\end{algorithm}
\begin{algorithm}[H]
\caption{Segment Search}
\label{alg:SegmentSearch}
\begin{algorithmic}[1]
\Function{Segment-Search}{$T_x,T_y,segment(=\overline{p_a p_c})$}
\State x-nodes are obtained by \Call{Interval-Search}{$T_x.root\,,\,[x_{p_a},x_{p_c}]$}
\State y-nodes are obtained by \Call{Interval-Search}{$T_y.root\,,\,[y_{p_a},y_{p_c}]$}
\For{x-node in x-nodes}
\For{y-node in y-nodes}
\If{x-node.segment is y-node.segment}
\State \text{output}(x-node.segment)
\EndIf
\EndFor
\EndFor
\EndFunction
\end{algorithmic}
\end{algorithm}
\subsubsection*{Step 3: Generating $P_{k+1}$ as $P_k \cup T_{k+1}$}
\begin{figure}[H]
    \centering
    \includegraphics[keepaspectratio, scale=0.3]{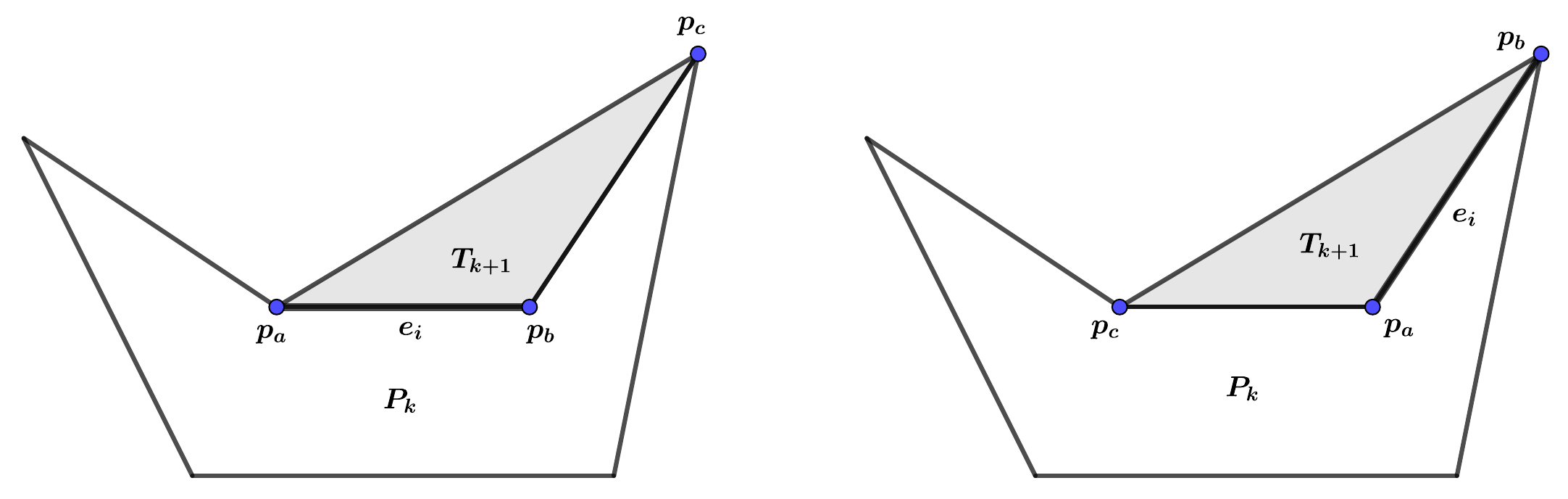}
    \caption{Generation method depends on edge $e_i$ connecting sequential points $B_{k}[i]$ and $B_{k}[i+1]$ in boundary sequence $B_k$.}
    \label{fig:3p2e}
\end{figure}

In Step 3, we consider the triangle $T_{k+1}$ adjacent to polygon $P_k$ with an edge $e_i$ that connects sequential points $B_{k}[i]$ and $B_{k}[i+1]$ in the boundary sequence $B_k$. Notably, this triangle should not be part of the existing polygon $P_k$.
Assuming the dataset formed by the union of $P_k$ and $T_{k+1}$ results in a valid triangulation in Step 2, the method of generating $B_k$ depends on the adjacency pattern. Let us consider the triangle $T_{k+1} = \Delta p_a p_b p_c$, where $p_a$ matches $B_k[i]$, $p_b$ aligns with $B_k[i+1]$, and $p_d$ corresponds to $B_k[i+2]$. Therefore, we can represent the boundary sequence $B_k$ as $[\ldots \,,\, p_a(=B_k[i]) \,,\, p_b(=B_k[i+1]) \,,\, p_d(=B_k[i+2]) \,,\, \ldots]$.
In the 2 points 1 edge shared scenario, we incorporate point $p_c$ between $p_a$ and $p_b$ in $B_k$ and update $i$ to $i+1$. This results in $B_k=[\ldots \,,\, p_a \,,\, p_c(=B_k[i]) \,,\, p_b(=B_k[i+1]) \,,\, \ldots]$. During the next iteration of Step 1, we look for triangles adjacent to edge $\overline{p_c p_b}$.
However, in the 3 points 2 edges shared scenario, the method for generating $B_k$ depends on the specific edge $e_i$, even for the same adjacent triangle $T_{k+1}$. For instance, when point $p_c$ is located on the side of $p_b$ (as shown in Fig.~\ref{fig:3p2e}, left), we exclude $p_b$ from $B_k$, yielding $B_k = [\ldots \,,\, p_a(=B_k[i]) \,,\, p_c(=B_k[i+1]) \,,\, \ldots]$. In the next iteration of Step 1, we search for triangles adjacent to edge $\overline{p_a p_c}$.
Alternatively, if point $p_c$ is situated on the side of $p_a$ (see Fig.~\ref{fig:3p2e}, right), we exclude $p_a$ from $B_k$ and decrement $i$ by 1. Hence, $B_k=[\ldots \,,\, p_c(=B_k[i]) \,,\, p_b(=B_k[i+1]) \,,\, \ldots]$. In the next iteration of Step 1, we hunt for triangles adjacent to edge $\overline{p_c p_b}$.
In the 3 point 1 edge shared scenario, we skip the adjacent triangle and refrain from merging it with the polygon $P_k$, thus updating $i$ to $i+1$. As a result, $B_k = [\ldots \,,\, p_a \,,\, p_b(=B_k[i]) \,,\, p_d(=B_k[i+1]) \,,\, \ldots]$. In the next iteration of Step 1, we seek triangles adjacent to edge $\overline{p_b p_d}$.
\newpage
\section{Verification of the Delaunay Property}\label{sec:veridp}
\begin{figure}[H]
    \centering
    \subfloat[Not locally Delaunay]{\label{fig:incircleNonDelaunay}\includegraphics[keepaspectratio, scale=0.4]{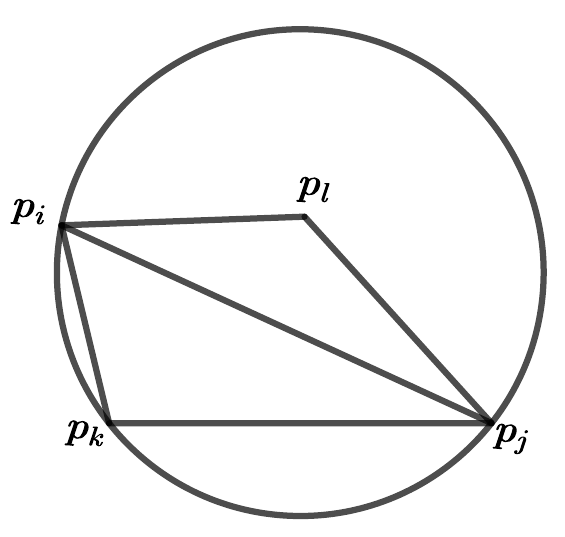}}
    \subfloat[Locally Delaunay]{\label{fig:incircleDelaunay}\includegraphics[keepaspectratio, scale=0.4]{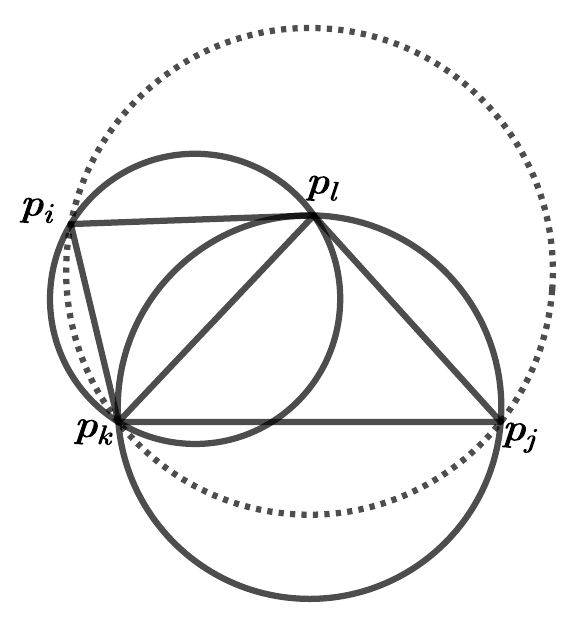}}
    \caption{In the context of the adjacent triangles $(p_i, p_j, p_k)$ and $(p_i, p_j, p_l)$, the diagram on the left violates the local Delaunay property, whereas that on the right satisfies it.}
    \label{fig:isDelaunay}
\end{figure}

If the correctness of triangulation is assured by the PSTV algorithm, then the Delaunay property can be verified through the following simple procedure. It must be emphasized that the correctness of the triangulation is a prerequisite for the verification of the Delaunay property.
We delineate the method for verifying the Delaunay property of a guaranteed-correct triangulation and the corrective measures to be taken if the Delaunay property is not satisfied. The Delaunay property, as referred to in this context, signifies the maximization of the minimum interior angle in the triangulation of a certain area that has been assured to be correct using the PSTV method. This property is equivalent to every edge in the triangulation being a Delaunay edge; we define a Delaunay edge in Definition \ref{def:Delaunay}. It is important to differentiate this concept of the Delaunay property, which is defined for triangulation, from the one defined for point sets.
\begin{dfn}\label{def:Delaunay}
An edge is said to exhibit the local Delaunay property if it satisfies either of the following conditions:
\begin{itemize}
\item The edge is part of only one triangle.
\item If the edge belongs to two triangles, the non-shared vertex of one triangle does not lie within the circumcircle of the other triangle.
\end{itemize}
An edge that possesses the local Delaunay property is referred to as a Delaunay edge.
\end{dfn}
A triangulation is said to have the Delaunay property when all its edges are Delaunay edges. This property corresponds to the maximization of the minimum interior angle in the triangulation, a notion that holds considerable importance in FEM applications. We now discuss a method for verifying the Delaunay property and the correctional measures to be adopted when it is not satisfied.
We can ascertain whether the edge $\overline{p_i p_j}$ shared between two triangles $(p_i, p_j, p_k)$ and $(p_i, p_j, p_l)$ exhibits the local Delaunay property by using the incircle test (see Fig.~\ref{fig:isDelaunay}). If an edge fails to exhibit the local Delaunay property, it can be altered to satisfy this property by removing edge $\overline{p_i p_j}$ and adding edge $\overline{p_k p_l}$, a process known as flipping. The edges that are intrinsic to the triangulation and are exempt from flipping are designated as constrained edges. A triangulation maximizes the minimum interior angle when all non-constrained edges have the local Delaunay property. If a triangulation fails to satisfy this property, the edges that lack the local Delaunay property are flipped. However, when flipping edge $\overline{p_i p_j}$, it is necessary to reassess whether edges $\overline{p_i p_k}$, $\overline{p_k p_j}$, $\overline{p_j p_l}$, and $\overline{p_l p_i}$ have the local Delaunay property, even if they initially did (see Algorithm \ref{alg:flip}).
\begin{algorithm}[H]
\caption{Flip}
\label{alg:flip}
\begin{algorithmic}[1]
\Function{Flip}{All edges $\mathbb{E}$}
\While{$Size\,of\,\mathbb{E} > 0$}
\State $\overline{p_i p_j}\,is\,one\,of\,\mathbb{E}$
\If{$Incircle(p_i, p_j, p_k, p_l) < 0$}
\State Flip $\overline{p_i p_j}$ (\,Replace $\overline{p_i p_j}$ with $\overline{p_k p_l}$\,)
\State Remove $\overline{p_i p_j}$ from $\mathbb{E}$
\For{each edge $\mathbf{of} \overline{p_i p_k} , \overline{p_k p_j} , \overline{p_j p_l} , \overline{p_l p_i}$}
\If{edge is not included in $\mathbb{E}$}
\State Append edge to $\mathbb{E}$
\EndIf
\EndFor
\EndIf
\EndWhile
\EndFunction
\end{algorithmic}
\end{algorithm}
\section{Numerical Verification Experiments}\label{sec:experiments}
\begin{figure}[H]
    \centering
    \begin{tabular}{cc}
        \begin{minipage}[t]{0.35\hsize}
            \centering
            \includegraphics[keepaspectratio, scale=0.33]{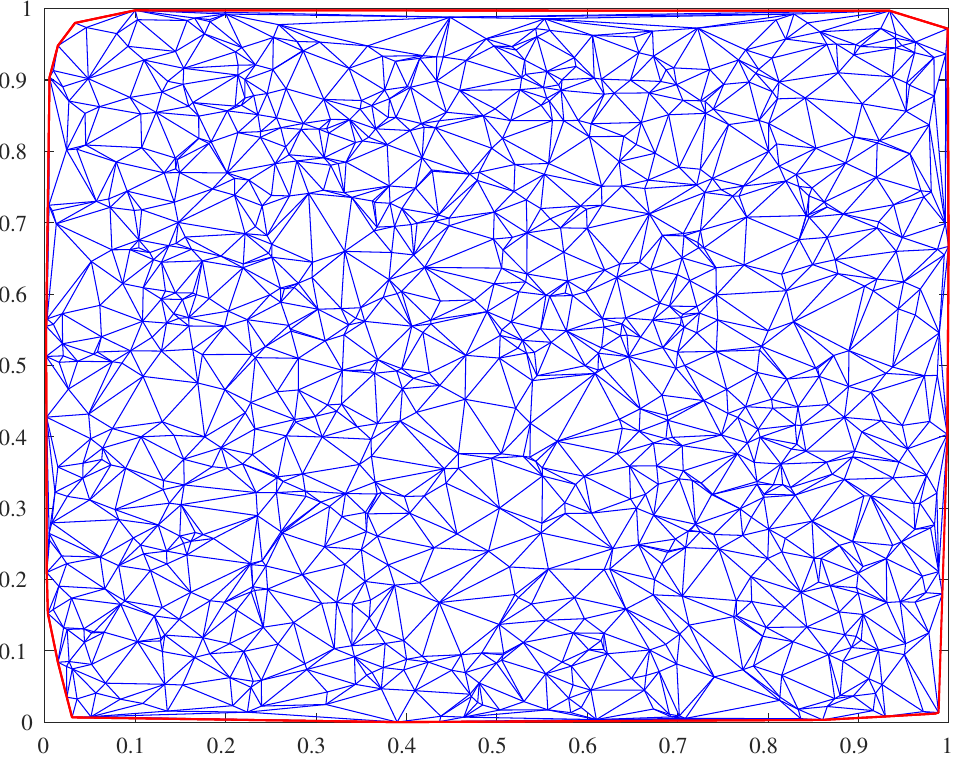}
            \subfloat{Uniform distribution}
            \label{fig:uniformData}
        \end{minipage}
        \begin{minipage}[t]{0.35\hsize}
            \centering
            \includegraphics[keepaspectratio, scale=0.33]{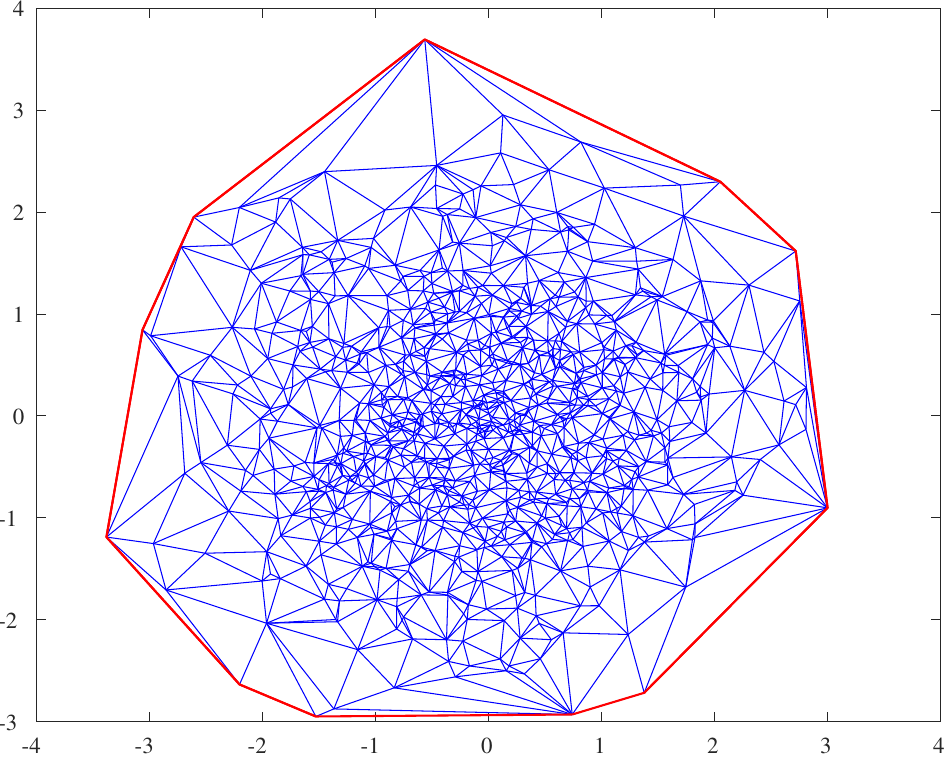}
            \subfloat{Normal distribution}
            \label{fig:normalData}
        \end{minipage}\\\\
        
        \begin{minipage}[t]{0.35\hsize}
            \centering
            \includegraphics[keepaspectratio, scale=0.33]{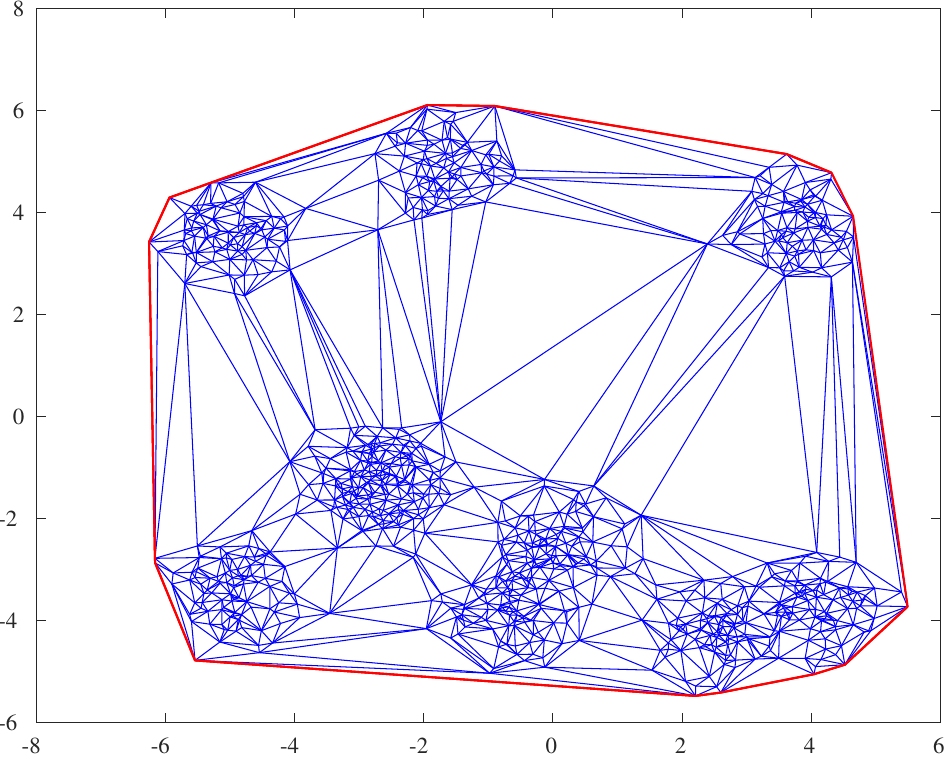}
            \subfloat{Cluster}
            \label{fig:clusterData}
        \end{minipage}
        \begin{minipage}[t]{0.35\hsize}
            \centering
            \includegraphics[keepaspectratio, scale=0.33]{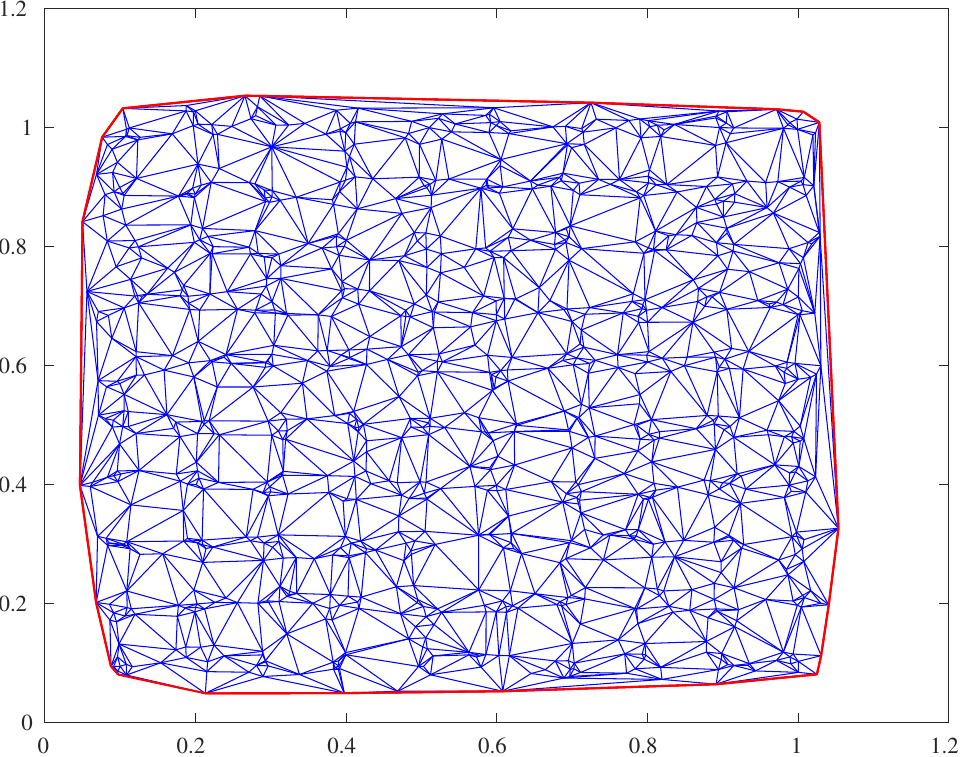}
            \subfloat{Grid}
            \label{fig:gridData}
        \end{minipage}
    \end{tabular}
    \caption{All datasets ({\rm N}=1000) generated in MATLAB R2022b. The cluster dataset (lower left) is divided into 10 clusters with nodes. The grid dataset (lower right) is distributed among 100 locations in a $10\times10$ grid.}
    \label{fig:dataset-patterns}
\end{figure}

\begin{figure}[ht]
    \centering
    \begin{tabular}{cc}
        \begin{minipage}[t]{0.3\hsize}
            \centering
            \includegraphics[keepaspectratio, scale=0.2]{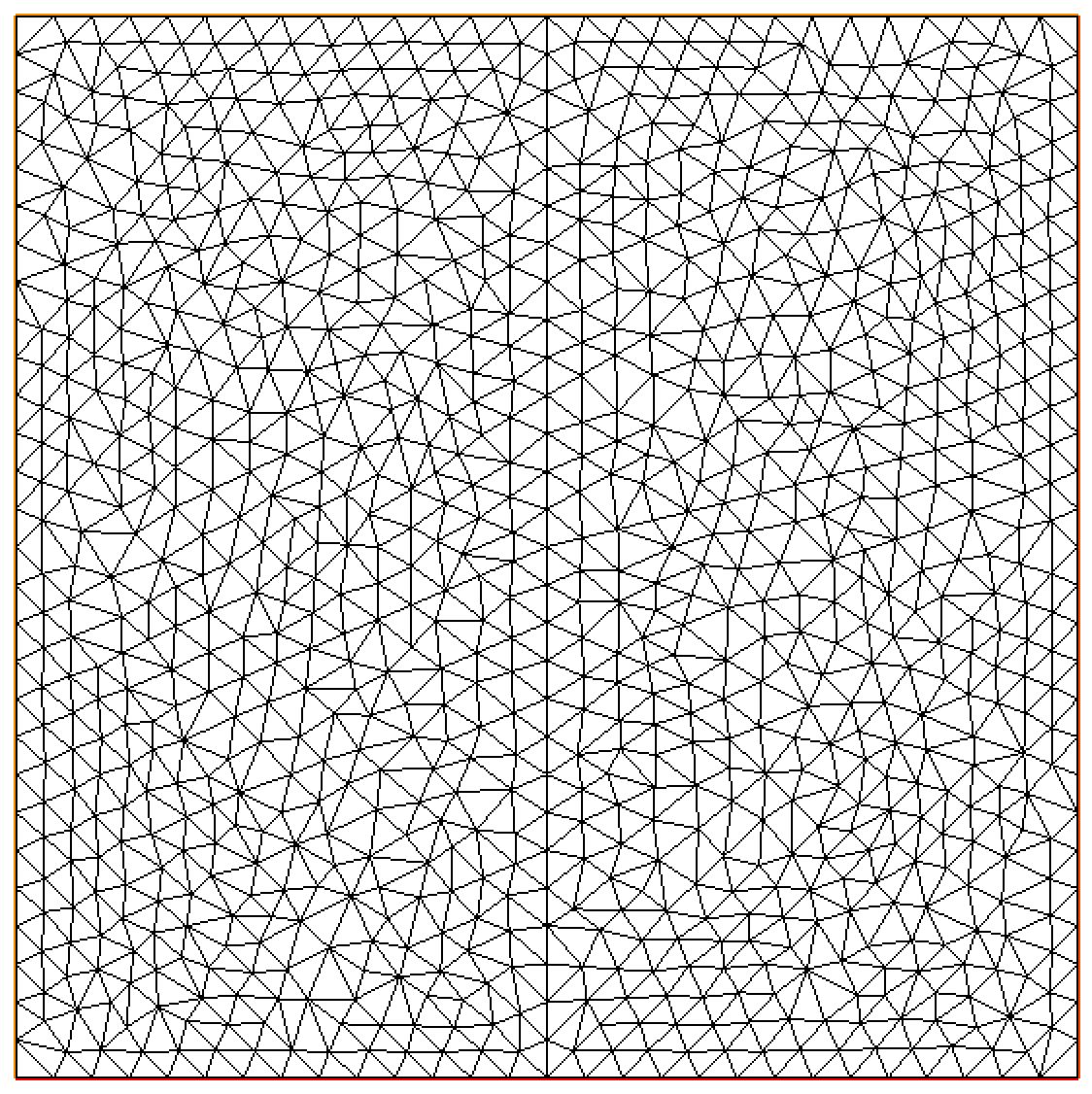}
            \subfloat{Square}
        \end{minipage}
        \begin{minipage}[t]{0.3\hsize}
            \centering
            \includegraphics[keepaspectratio, scale=0.2]{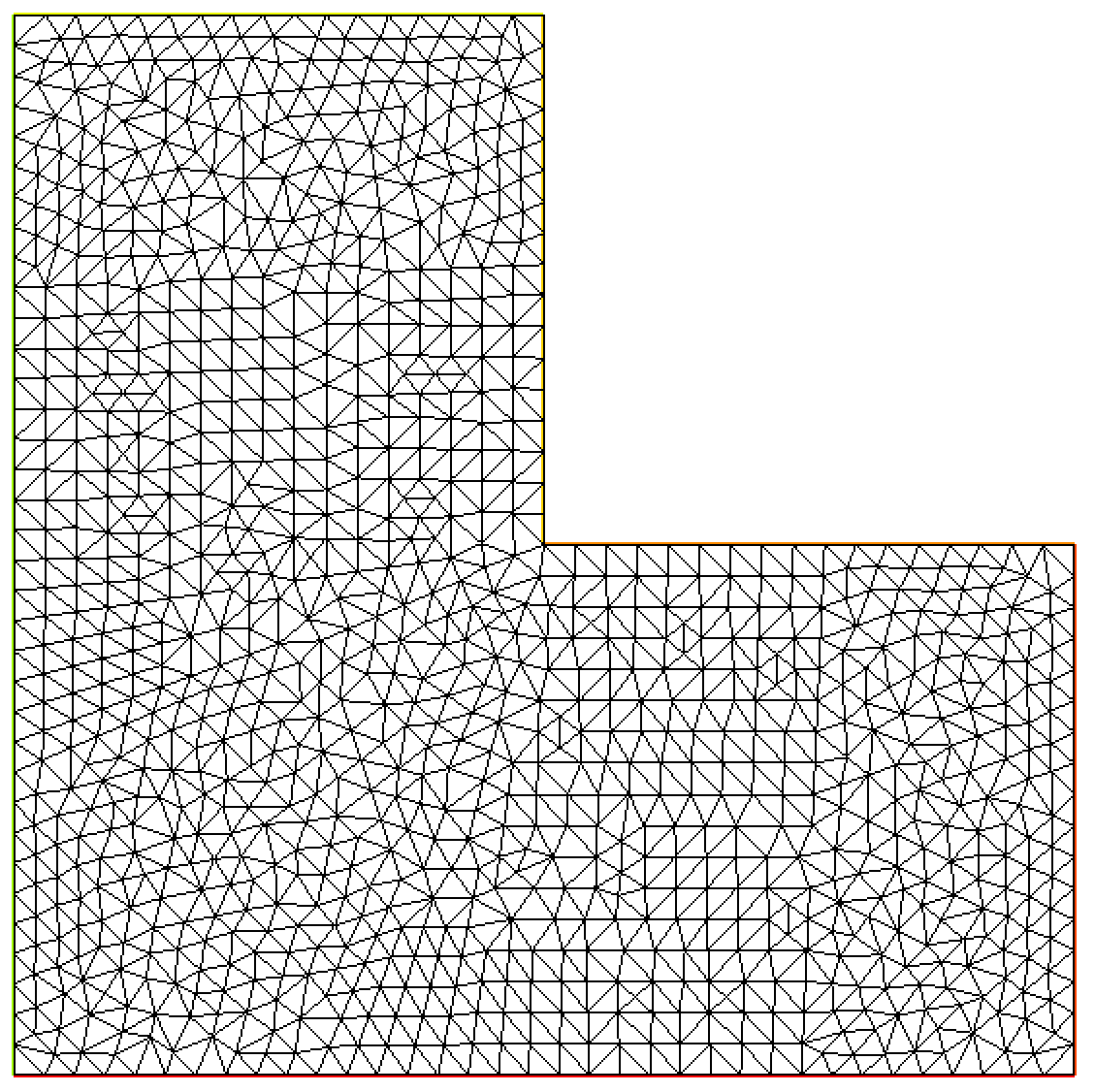}
            \subfloat{L-shaped}
        \end{minipage}
    \end{tabular}
    \caption{These datasets were generated in FreeFEM++ v4.9. The calculations were performed using the ``buildmesh'' function with the specified outer boundary.}
    \label{fig:square-L}
\end{figure}

A series of numerical experiments were conducted on a computer with a 4.70 GHz AMD Ryzen 9 7900X 12-core processor, 128 GB RAM, the Ubuntu 22.04 operating system, GMP Version 6.2.1, and GCC Version 9.4.0.
Regarding the interval tree process in the PSTV algorithm, the program code is taken from \cite{ebbeke2022intervaltree}.
In this algorithm, it is crucial to discern the origin of intervals held in the X- and Y-interval trees. 
Therefore, the program code was modified to ensure clarity regarding the source segments of these intervals.
We created a dataset using a set of $N$ nodes distributed in four different patterns.
The first pattern involves a set of nodes, denoted as $P$, distributed uniformly over the domain $\Omega = (0,1)^2$.
The second pattern involves a set of nodes, also denoted as $P$, distributed according to a standard normal distribution.
The third pattern involves a set of $10$ distinct random nodes, each located at a different center within the domain $\Omega = (-5,5)^2$. Around each center, there are $N/10$ nodes distributed according to a normal distribution with a standard deviation of $0.5$.
The fourth pattern involves $100$ nodes, each located at a center that satisfies the condition $\{(x, y) | x \in \mathbb{Z}, y \in \mathbb{Z}, 1 \leq x \leq 10, 1 \leq y \leq 10\}$. Around each center, there are $N/100$ nodes distributed according to a normal distribution $\mathcal{N}(0,0.04)$.
These four patterns sequentially represent the uniform, normal, cluster, and grid configurations in Tables \ref{tab:num-of-triangles} and \ref{tab:num-of-edges}.
The results of drawing each pattern with 1000 nodes are shown in Fig.~\ref{fig:dataset-patterns}.
We used the ``delaunayTriangulation'' function in MATLAB to obtain the convex hull of set $P$ and the set of triangles constituting its Delaunay triangulation.
We now have the necessary datasets from Section \ref{sec:prep}, including the set of vertices, the set of triangles, and the sequence of boundary vertices.
We performed numerical experiments to verify whether each dataset is a valid triangulation and, if so, whether all the edges that constitute the triangulation satisfy the local Delaunay property.
The time taken to output the Delaunay triangulation in MATLAB, the computation time for the PSTV algorithm to verify the correctness of the triangulation, and the computation time for determining whether the triangulation satisfies the minimum interior angle maximization for correct cases are presented in Tables \ref{tab:uniform}, \ref{tab:normal}, \ref{tab:cluster}, and \ref{tab:grid}.
\begin{table}[H]
\centering
\caption{Number of triangles.}
\label{tab:num-of-triangles}
\begin{tabular}{l|cccccccc}
\hline
N       & 1000 & 2000 & 4000 & 8000  & 16000 & 32000 & 64000  & 128000 \\ \hline
uniform & 1977 & 3977 & 7977 & 15974 & 31973 & 63972 & 127968 & 255969 \\ \hline
normal  & 1987 & 3987 & 7984 & 15982 & 31984 & 63983 & 127976 & 255981 \\ \hline
cluster & 1983 & 3984 & 7981 & 15981 & 31984 & 63983 & 127976 & 255979 \\ \hline
grid    & 1977 & 3975 & 7980 & 15978 & 31981 & 63976 & 127975 & 255981 \\ \hline
\end{tabular}
\end{table}
\begin{table}[H]
\centering
\caption{Number of edges.}
\label{tab:num-of-edges}
\begin{tabular}{l|cccccccc}
\hline
N       & 1000 & 2000 & 4000   & 8000  & 16000 & 32000 & 64000  & 128000 \\ \hline
uniform & 2976 & 5976 & 11976  & 23973 & 47972 & 95971 & 191967 & 383968 \\ \hline
normal  & 2986 & 5986 & 11983  & 23981 & 47983 & 95982 & 191975 & 383980 \\ \hline
cluster & 2982 & 5983 & 119800 & 23980 & 47983 & 95983 & 191976 & 383978 \\ \hline
grid    & 2976 & 5974 & 11979  & 23977 & 47980 & 95975 & 191974 & 383980 \\ \hline
\end{tabular}
\end{table}
\begin{table}[H]
\centering
\caption{Computation times for uniform distribution.}
\label{tab:uniform}
\small
\begin{tabular}{l|cccccccc}
\hline
N    & 1000   & 2000   & 4000   & 8000   & 16000  & 32000  & 64000  & 128000 \\ \hline
A    & 1.1E-2 & 3.3E-2 & 8.8E-2 & 2.6E-1 & 7.9E-1 & 2.4    & 7.6    & 2.4E1  \\ \hline
B    & 2.2E-2 & 4.7E-2 & 1.0E-1 & 2.2E-1 & 5.0E-1 & 1.0    & 2.4    & 5.7    \\ \hline
C    & 1.48   & 1.53   & 1.58   & 1.82   & 2.13   & 2.77   & 3.52   & 5.62   \\ \hline
\end{tabular}
\begin{itemize}
\item[A.] Verification time of triangulation [s]
\item[B.] Verification time of minimum interior angle maximization [s]
\item[C.] Time required to compute an approximate solution in MATLAB [s]
\end{itemize}
\end{table}
\begin{table}[H]
\centering
\caption{Computation times for normal distribution.}
\label{tab:normal}
\small
\begin{tabular}{l|cccccccc}
\hline
N    & 1000   & 2000   & 4000   & 8000   & 16000  & 32000  & 64000  & 128000 \\ \hline
A    & 1.2E-2 & 3.5E-2 & 9.4E-2 & 2.7E-1 & 8.1E-1 & 2.6    & 8.1    & 2.6E1  \\ \hline
B    & 2.1E-2 & 4.3E-2 & 8.6E-2 & 1.7E-1 & 3.6E-1 & 7.6E-1 & 1.5    & 3.1    \\ \hline
C    & 1.49   & 1.53   & 1.61   & 1.71   & 2.42   & 2.78   & 3.61   & 5.75   \\ \hline
\end{tabular}
\end{table}
\begin{table}[H]
\centering
\caption{Computation times for cluster.}
\label{tab:cluster}
\small
\begin{tabular}{l|cccccccc}
\hline
N    & 1000   & 2000   & 4000   & 8000   & 16000  & 32000  & 64000  & 128000 \\ \hline
A    & 1.2E-2 & 3.2E-2 & 9.2E-2 & 2.8E-1 & 8.4E-1 & 2.7    & 8.9    & 2.8E1  \\ \hline
B    & 2.1E-2 & 4.5E-2 & 9.0E-2 & 1.8E-1 & 3.7E-1 & 7.8E-1 & 1.6    & 3.4    \\ \hline
C    & 1.49   & 1.53   & 1.60   & 1.74   & 1.97   & 2.45   & 3.42   & 5.26   \\ \hline
\end{tabular}
\end{table}
\begin{table}[H]
\centering
\caption{Computation times for grid.}
\label{tab:grid}
\small
\begin{tabular}{l|cccccccc}
\hline
N    & 1000   & 2000   & 4000   & 8000   & 16000  & 32000  & 64000  & 128000 \\ \hline
A    & 1.1E-2 & 3.3E-2 & 9.9E-2 & 3.0E-1 & 8.8E-1 & 2.9    & 9.0    & 2.8E1  \\ \hline
B    & 2.3E-2 & 4.5E-2 & 9.6E-2 & 1.9E-1 & 4.1E-1 & 8.5E-1 & 1.7    & 3.7    \\ \hline
C    & 1.51   & 1.54   & 1.61   & 1.76   & 2.03   & 2.66   & 3.46   & 5.24   \\ \hline
\end{tabular}
\end{table}
FEM researchers often employ FreeFEM++ \cite{MR3043640} to acquire meshes for their domains. In FreeFEM++, it is possible to determine the mesh for a domain by considering the convex hull of its nodes, or by explicitly specifying the outer boundary to create the mesh. This means that meshes can be generated for nonconvex domains.
Experiments were conducted using both a square domain, denoted as $\Omega = (-1,1)^2$, and an L-shaped domain, denoted as $\Omega = (-1,1)^2 \setminus (0,1)^2$. Similar experiments were carried out for ten different datasets to ascertain whether they satisfy the minimum internal angle maximization criterion. In cases where this criterion was not met, an examination was conducted to determine the extent of edges that did not have the Delaunay property.
Each dataset comprises vertices that fall within $\pm 5\%$ of $N$.
The square domain (see Fig.~\ref{fig:square-L}) corresponds to a triangular mesh dataset generated by dividing the perimeter of the square into segments of $N_b$, $N_b+1$, or $N_b-1$ such that the node count closely approximates $N$.
For the L-shaped domain (see Fig.~\ref{fig:square-L}), the first step involves partitioning the perimeter along the x and y axes, resulting in eight segments. Subsequently, these segments are further divided using $N_b$, $N_b+1$, or $N_b-1$ to obtain a triangular mesh dataset with a node count that approximates $N$.
We conducted numerical experiments using the 10 datasets generated as described above (see Tables \ref{tab:square} and \ref{tab:L}).
\begin{table}[H]
\centering
\caption{Computation times and probabilities of the triangulation not satisfying the minimum interior angle maximization (square domain).}
\label{tab:square}
\begin{tabular}{l|ccccccccc}
\hline
$N$ & 1000 & 2000 & 4000 & 8000 & 16000 & 32000 & 64000 & 128000 \\ \hline
$N_b$ & 28 & 40 & 58 & 81 & 116 & 164 & 232 & 327 \\ \hline
$N_{min}$ & 975 & 1973 & 4079 & 7937 & 16016 & 32166 & 64173 & 127906 \\ \hline
$N_{max}$ & 1019 & 2030 & 4190 & 8129 & 16544 & 32976 & 66080 & 130904 \\ \hline
$N_{ave}$ & 998.2 & 1998.4 & 4147 & 8052 & 16415.3 & 32774.8 & 65495.3 & 130275 \\ \hline
$E_{min}$ & 2704 & 5602 & 11774 & 23164 & 47121 & 95187 & 190664 & 381103\\ \hline
$E_{max}$ & 2832 & 5769 & 12207 & 23738 & 48703 & 97615 & 196383 & 290095\\ \hline
$E_{ave}$ & 2771.6 & 5676.2 & 11978 & 23509 & 48318.9 & 97013.4 & 194631 & 388210\\ \hline
$T_{min}$ & 1837 & 3785 & 7924 & 15548 & 31566 & 63674 & 127416 & 254502\\ \hline
$T_{max}$ & 1923 & 3897 & 8213 & 15931 & 32621 & 65293 & 131229 & 260497\\ \hline
$T_{ave}$ & 1882.4 & 3834.8 & 8060 & 15778 & 32364.6 & 64891.6 & 130061 & 259240\\ \hline
$A$ & 1.0E-2 & 2.9E-2 & 8.8E-2 & 2.4E-1 & 7.2E-1 & 2.3 & 7.1 & 2.2E1\\ \hline
$B$ & 1.8E-2 & 3.7E-2 & 7.9E-2 & 1.5E-1 & 3.3E-1 & 6.9E-1 & 1.4 & 2.8\\ \hline
$C$ & 3.3E-1 & 6.5E-1 & 6.8E-1 & 7.3E-1 & 8.3E-1 & 1.1 & 1.5 & 2.5\\ \hline
$D$ & 0 & 2 & 2 & 13.5 & 2 & 2 & 12 & 36.8\\ \hline
$F$ & 0 & 10 & 10 & 90 & 10 & 20 & 10 & 90\\ \hline
\end{tabular}
\begin{itemize}
\item[$N_{min}$.] Minimum number of nodes among all datasets
\item[$N_{max}$.] Maximum number of nodes among all datasets
\item[$N_{ave}$.] Average number of nodes across all datasets
\item[$E_{min}$.] Minimum number of edges among all datasets
\item[$E_{max}$.] Maximum number of edges among all datasets
\item[$E_{ave}$.] Average number of edges across all datasets
\item[$T_{min}$.] Minimum number of triangles among all datasets
\item[$T_{max}$.] Maximum number of triangles among all datasets
\item[$T_{ave}$.] Average number of triangles across all datasets
\item[D.] Average number of non-Delaunay edges in datasets that do not satisfy the minimum internal angle maximization condition
\item[F.] Percentage of datasets that do not satisfy the minimum internal angle maximization condition
\end{itemize}
\end{table}
\begin{table}[H]
\centering
\caption{Computation times and probabilities of the triangulation not satisfying the minimum interior angle maximization (L-shaped domain).}
\label{tab:L}
\begin{tabular}{l|ccccccccc}
\hline
$N$ & 1000 & 2000 & 4000 & 8000 & 16000 & 32000 & 64000 & 128000 \\ \hline
$N_b$ & 16 & 24 & 34 & 48 & 67 & 96 & 137 & 194 \\ \hline
$N_{min}$ & 951 & 2026 & 4003 & 7882 & 15423 & 31361 & 63860 & 127993 \\ \hline
$N_{max}$ & 979 & 2097 & 4168 & 8210 & 16067 & 32445 & 65006 & 130070 \\ \hline
$N_{ave}$ & 958 & 2065.4 & 4090.7 & 8069.6 & 15707.7 & 31752.3 & 64466.3 & 128869 \\ \hline
$E_{min}$ & 2538 & 5699 & 11470 & 22883 & 45198 & 92552 & 189393 & 380878\\ \hline
$E_{max}$ & 2678 & 5947 & 11961 & 23863 & 47130 & 95802 & 192827 & 387107\\ \hline
$E_{ave}$ & 2616.4 & 5814.6 & 11730.5 & 23443.2 & 46049.5 & 93723.3 & 191209 & 383506\\ \hline
$T_{min}$ & 1731 & 3859 & 7733 & 15739 & 30307 & 61953 & 126623 & 254432\\ \hline
$T_{max}$ & 1825 & 4025 & 8061 & 16033 & 31595 & 64120 & 128913 & 258585\\ \hline
$T_{ave}$ & 1783.7 & 3936.5 & 7907.1 & 15752.9 & 30875.1 & 62734.3 & 127834 & 256184\\ \hline
$A$ & 8.8E-3 & 3.0E-2 & 8.3E-2 & 2.2E-1 & 5.9E-1 & 1.9 & 6.6 & 2.1E1\\ \hline
$B$ & 1.6E-2 & 3.6E-2 & 7.7E-2 & 1.5E-1 & 3.1E-1 & 6.6E-1 & 1.4 & 2.8\\ \hline
$C$ & 3.1E-1 & 6.5E-1 & 6.7E-1 & 7.2E-1 & 8.2E-1 & 1.0 & 1.6 & 2.5\\ \hline
$D$ & 2 & 8.5 & 13.7 & 18.2 & 10 & 42.7 & 51.6 & 74.6\\ \hline
$F$ & 10 & 70 & 70 & 80 & 10 & 80 & 100 & 100\\ \hline
\end{tabular}
\end{table}
The current application does not handle datasets with overlapping edges. For instance, when generating a domain that resembles a circular region with narrow incisions using FreeFEM++, the datasets may have edges that overlap near the center. Such datasets will produce errors, leading to immediate termination of the computation (see Fig.~\ref{fig:center-thread}).
\begin{figure}[H]
    \centering
    \begin{tabular}{cc}
        \begin{minipage}[t]{0.45\hsize}
            \centering
            \includegraphics[keepaspectratio, scale=0.15]{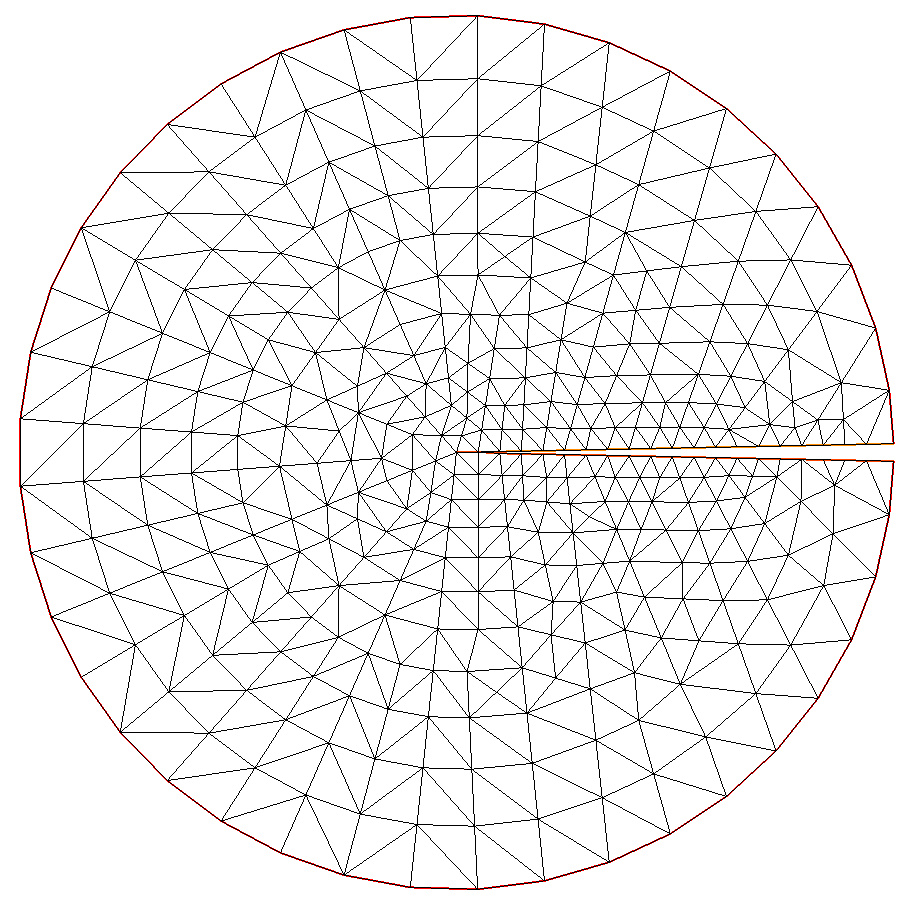}
            \label{fig:incision1}
        \end{minipage}
        \begin{minipage}[t]{0.45\hsize}
            \centering
            \includegraphics[keepaspectratio, scale=0.28]{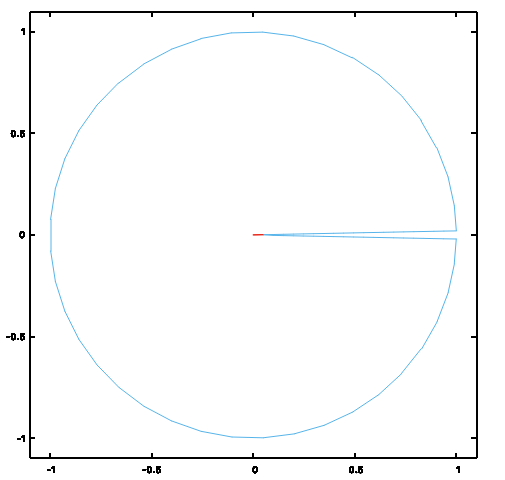}
            \label{fig:incision2}
        \end{minipage}
    \end{tabular}
    \caption{Dataset generated using FreeFEM++ v4.9. In the triangular mesh, the edge forms a thread in the red central region, as depicted in the right figure.}
    \label{fig:center-thread}
\end{figure}
\section{Conclusion}\label{sec:conc}
When using existing software to generate a triangulation, there is a chance that the output will be inaccurate. In the case of generating Delaunay triangulations, there is a possibility that the resulting dataset does not satisfy the minimum angle maximization criterion. 
Hence, it is essential to perform {\emph a posteriori} validation to ascertain the correctness of the computational results.
The PSTV algorithm provides {\emph a posteriori} verification to assess the accuracy of the triangulation.
If the triangulation is deemed correct, the algorithm subsequently verifies whether it satisfies the minimum interior angle maximization criterion.
In instances where this criterion is not met, the algorithm generates an adjusted dataset that satisfies the criterion.
{\emph A posteriori} validation of the dataset is independent of software, enabling the accuracy of the triangulation to be confirmed.
Moreover, if the region is simply connected, it is possible to validate the triangulation for any arbitrary region.
The PSTV computation time is similar to the output time of the approximate calculations, unless the mesh is exceptionally fine.
We have developed a web application to facilitate the easy use of the PSTV algorithm.
Users simply input the triangulation dataset, and the web application verifies its correctness and compliance with the minimum interior angle maximization criterion.
The developed software is accessible at the following URL: \url{https://github.com/uchunanora/2d-triangulation-validator}

\appendix
\section{Floating-point Filter}
Let $\mathbb{F}$ be a set of binary floating-point numbers, as defined in IEEE 754.
Let $\mathtt{fl}(\cdot)$ denote the result computed by floating-point arithmetic, where the rounding mode is rounding to nearest (ties to even).
Let $u$, $u_n$, and $u_s$ be the roundoff unit, minimum positive normalized floating-point number, and minimum positive floating-point number, respectively.
For binary64, $(u, u_n, u_s) = (2^{-53}, 2^{-1022}, 2^{-1074})$.
For $a, b \in \mathbb{F}$, the IEEE 754 standard specifies
\begin{equation}
\mathtt{fl}(a \pm b) = (a+b)(1 + \delta), \quad |\delta | \le u, 
\label{eq:error_sum}
\end{equation}
\begin{equation}
a + b = \mathtt{fl}(a \pm b)(1 + \delta), \quad |\delta | \le u, 
\label{eq:error_sum2}
\end{equation}
and
\begin{equation}
\mathtt{fl}(a \cdot b) = (a \cdot b)(1 + \delta)+\eta, \quad |\delta | \le u, \quad |\eta| \le \frac{1}{2}u_s, \quad \delta \cdot \eta = 0, 
\label{eq:error_mul}
\end{equation}
where we assume that overflow does not occur in $\mathtt{fl}(\cdot)$.
Let $u_i := (1+u)^i$.
These inequalities are applied to the derivation of the floating-point filter, for example, 
\[
|a + b| \le (1+u) \mathtt{fl}(|a+b|) \le (1+u) \mathtt{fl}(|a|+|b|) = u_i \mathtt{fl}(|a|+|b|)
\]
and
\[
|a \cdot b| \le (1+u) \mathtt{fl}(|a \cdot b|) + \frac{1}{2}u_s = u_i \mathtt{fl}(|a \cdot b|) + \frac{1}{2}u_s.
\]
First, we analyze the rounding errors for $\alpha_a$ in Algorithm~\ref{alg:verified-ict}:
\[
\alpha_a := \mathtt{fl}\left(\ \left( adx^2+ady^2 \right) (bdx \cdot cdy-bdy \cdot cdx)\ \right), 
\]
where 
\begin{align*}
adx & = \mathtt{fl}(x_a - x_d), \quad bdx = \mathtt{fl}(x_b - x_d), \quad cdx = \mathtt{fl}(x_c - x_d), \\
ady & = \mathtt{fl}(y_a - y_d), \quad bdy = \mathtt{fl}(y_b - y_d), \quad cdy = \mathtt{fl}(y_c - y_d).
\end{align*}
We use $\delta_i$ and $\eta_i$ satisfying $|\delta_i| \le u$ and $|\eta_i| \le u_s/2$ for all $i$.
We consider the rounding errors for $A, B \in \mathbb{F}$: 
\begin{align*}
\alpha_a = \mathtt{fl}(AB) = AB(1+\delta_1) + \eta_1, \\
A:=\mathtt{fl}\left(adx^2+ady^2 \right), \quad 
B:=\mathtt{fl}(bdx \cdot cdy-bdy \cdot cdx).
\end{align*}
Because $A$ and $B$ have the same structure, we focus on $A$ for the rounding error analysis:
\begin{align*}
    A := & \mathtt{fl}(\ adx^2+ady^2\ ) 
    = \left(\ \mathtt{fl}(adx^2)+\mathtt{fl}(ady^2) \ \right)(1+\delta_2) \\
    = & (\ adx^2(1+\delta_3) + \eta_3 +ady^2(1+\delta_4)+\eta_4\ ) (1+\delta_2)\\
    = & \left( \ (x_a - x_d)^2 (1 + \delta_5)^2(1+\delta_3) + \eta_3
      + (y_a - y_d)^2(1+\delta_6)^2 (1+\delta_4) + \eta_4 \ \right)(1+\delta_2).
\end{align*}
Similarly, we have
\begin{align*}
B = ( \ (b_x - d_x)(c_y - d_y) (1 + \delta_8)^2(1+\delta_9) + \eta_8 - (b_y-d_y)(c_x-d_x) (1+\delta_{10})^2 (1+\delta_{11}) + \eta_{10} \ )(1+\delta_7).
\end{align*}
Let the exact values of $\alpha_A$, $\alpha_B$, and $\alpha_C$ be $\alpha_A'$, $\alpha_B'$, and $\alpha_C'$, respectively, and let $(1+u)^9-1 = u_9 - 1$ be $\theta$.
Then, we derive
\begin{align}
|\alpha_a - \alpha_a'| \le & |AB (1+\delta_1) + \eta_1 - \alpha_a'| \label{eq:diff01}\\
& \left( u_9-1 \right) \left( \ (x_a - x_d)^2 + (y_a - y_d)^2 \ \right) \left( \ |(x_b-x_d)(y_c - y_d)| + |(y_b- y_d)(x_c- x_d)| \ \right) \label{eq:diff02} \\
   & \quad + u_s \cdot u_6 \left( \ |(x_b-x_d)(y_c - y_d)| + |(y_b- y_d)(x_c- x_d)| \ \right) \label{eq:diff03} \\
   & \quad + u_s \cdot u_6 \left( \ (x_a - x_d)^2 + (y_a - y_d)^2 \ \right)  + 4\eta^2 \cdot u_3 + \frac{1}{2}u_s \label{eq:diff04} \\
   \le &  \theta \left( \ (x_a - x_d)^2 + (y_a - y_d)^2 \ \right) \left( \ |(x_b-x_d)(y_c - y_d)| + |(y_b- y_d)(x_c- x_d)| \ \right) \nonumber \\
   & + u_s \cdot u_8 \left( \ |bdx \cdot cdy | + | bdy \cdot cdx | \ \right) + u_s \cdot u_8 \left( adx^2 + ady^2 \right) + u_s^2 \cdot u_3 + \frac{1}{2}u_s \nonumber \\
      < &  \theta \cdot u_2\left( \ adx^2 + ady^2 \ \right) u_2 \left( \ |bdx \cdot cdy | + |bdy \cdot cdx| \ \right) \nonumber \\
   & + u_s \cdot u_9 \left( \ |\mathtt{fl}(bdx \cdot cdy)| + |\mathtt{fl}(bdy \cdot cdx)| + u_s \ \right) \nonumber \\
   & + u_s \cdot u_9 \left( \mathtt{fl}(adx^2) + \mathtt{fl}(ady^2) + u_s \right) + u_s^2 \cdot u_3 + \frac{1}{2}u_s \nonumber\\
      < &  \theta \cdot u_4 \left( \alpha_{a1} + u_s\right) * u_4 ( \alpha_{a2'} + u_s) \nonumber \\
   & + u_s \cdot u_{10} \left( \alpha_{a2'} + u_s \right) + u_s \cdot u_{10} \left( \alpha_{a1} + u_s \right) + u_s^2 \cdot u_3 + \frac{1}{2}u_s \nonumber \\
      \le & \theta \cdot u_9 \alpha_{a'} + \frac{1}{2}u_s  +
      u_s \theta \cdot u_8 ( \alpha_{a1} + \alpha_{a2'} ) + u_s^2 \theta \cdot u_8 \nonumber \\
   & + u_s \cdot u_{10} \left( \alpha_{a2'} + u_s \right) + u_s \cdot u_{10} \left( \alpha_{a1} + u_s \right) + u_s^2 \cdot u_3 + \frac{1}{2}u_s \nonumber\\
         \le & \theta \cdot u_9 \alpha_{a'}  +
       \theta u_s \cdot u_9 (\mathtt{fl}(\alpha_{a1} + \alpha_{a2'})) + u_s \cdot u_{11} \mathtt{fl}\left(\alpha_{a2'} + \alpha_{a1} \right) + 4 u_s^2 \cdot u_{10} + u_s \nonumber \\
    \le & \theta \cdot u_9 \cdot \alpha_{a'} + 2u_s \cdot u_{11} \mathtt{fl}\left(\alpha_{a2'} + \alpha_{a1} \right) + 4 u_s^2 \cdot u_{10} + u_s. \nonumber
\end{align}
From \eqref{eq:diff01}--\eqref{eq:diff02}, \eqref{eq:diff03}, and \eqref{eq:diff04}, despite the complicated computations, we derive a simple strategy.
We expand \eqref{eq:diff01} and take an upper bound using $|\delta_i| \le u$.
We can obtain similar results for $\alpha_b$ and $\alpha_c$, 
such that
\[
\alpha_{a}' - \alpha_a = \delta_1', \quad \alpha_{b}' - \alpha_b = \delta_2', \quad \alpha_{c}' - \alpha_c = \delta_3',
\]
where
\begin{align*}
|\delta_1'| & \le \theta \cdot u_9 \alpha_{a'} + 2u_s \cdot u_{11} \mathtt{fl}\left(\alpha_{a2'} + \alpha_{a1} \right) + 4 u_s^2 \cdot u_{10} + u_s, \\
|\delta_2'| & \le \theta \cdot u_9 \alpha_{b'} + 2u_s \cdot u_{11} \mathtt{fl}\left(\alpha_{b2'} + \alpha_{b1} \right) + 4 u_s^2 \cdot u_{10} + u_s, \\
|\delta_3'| & \le \theta \cdot u_9 \alpha_{c'} + 2u_s \cdot u_{11} \mathtt{fl}\left(\alpha_{c2'} + \alpha_{c1} \right) + 4 u_s^2 \cdot u_{10} + u_s.
\end{align*}
Here, we have assumed that $\theta < 1$, as is naturally satisfied for binary16, 32, 64, and 128 in the IEEE 754 standard.
Now, we have
\[
\alpha_{a}' + \alpha_{b}' + \alpha_{c}' = \alpha_a + \alpha_b + \alpha_c + \delta_1' + \delta_2' + \delta_3'
\]
and 
\begin{align*}
\alpha_{a}' + \alpha_{b}' + \alpha_{c}' & = \mathtt{fl}(\alpha_a + \alpha_b) + \delta_4'\mathtt{fl}(\alpha_a + \alpha_b) + \alpha_c + \delta_1' + \delta_2' + \delta_3' \\
 & = (1+\delta_5')\mathtt{fl}((\alpha_a + \alpha_b)+\alpha_c) + \delta_4' \mathtt{fl}(\alpha_a + \alpha_b) + \delta_1' + \delta_2' + \delta_3'.
 \end{align*}
Therefore, if
\[
(1 - u)|\mathtt{fl}((\alpha_a + \alpha_b)+\alpha_c)| > u |\mathtt{fl}(\alpha_a + \alpha_b)| + |\delta_1'| + |\delta_2'| + |\delta_3'|, 
\]
namely, 
\begin{equation}
|\mathtt{fl}((\alpha_a + \alpha_b)+\alpha_c)| > \frac{u |\mathtt{fl}(\alpha_a + \alpha_b)| + |\delta_1'| + |\delta_2'| + |\delta_3'|}{1-u}
\label{eq:th_bound}    
\end{equation}
is satisfied, the sign of the computed determinant is correct.
Let 
\[
\omega := \mathtt{fl}(((\alpha_{a2'} + \alpha_{a1}) + (\alpha_{b2'}+ \alpha_{b1} )) + (\alpha_{c2'} + \alpha_{c1} )), \quad \gamma := \mathtt{fl}((\alpha_{a'}+\alpha_{b'})+\alpha_{c'}).
\]
The upper bound of $|\delta_1'| + |\delta_2'| + |\delta_3'|$ is given by
\begin{align*}
&|\delta_1'| + |\delta_2'| + |\delta_3'| \\
\le& \theta \cdot u_9  (\alpha_{a'}+\alpha_{b'}+\alpha_{c'}) 
+ 2u_s \cdot u_{11} \{ \mathtt{fl}\left(\alpha_{a2'} + \alpha_{a1} \right) + \mathtt{fl}\left(\alpha_{b2'} + \alpha_{b1} \right) + \mathtt{fl}\left(\alpha_{c2'} + \alpha_{c1} \right) \}
+ 12 u_s^2 \cdot u_{10} + 3u_s\\
\le& \theta \cdot u_{11} \cdot \gamma 
+ 2u_s \cdot u_{13} \omega  + 12 u_s^2 \cdot u_{10} + 3u_s
\end{align*}
Hence, we have
\begin{align}
& \frac{u |\mathtt{fl}(\alpha_a + \alpha_b)| + |\delta_1'| + |\delta_2'| + |\delta_3'|}{1-u} \nonumber \\
\le& 
\frac{\left( \theta \cdot u_{11} + u\right) \gamma}{1-u} \frac{u_2}{u_2} \nonumber + \frac{2u_s \cdot u_{13}}{1-u} \omega \nonumber + \frac{12 u_s^2 \cdot u_{10}}{1-u} + \frac{3}{1-u} u_s \nonumber \\
\le& \frac{(\theta \cdot u_{11} + u) u_2}{1-u} \frac{\gamma}{u_2}  + \frac{2u_s \cdot u_{13}}{1-u} \omega + 4u_s 
\le \frac{( \theta \cdot u_{11} + u) u_2 }{(1-u)u_2} \gamma
+ \frac{4u_s \cdot u_{16}}{(1-u)u_2}  (\omega  + 1) 
\label{eq:second}
\end{align}
We compute the upper bounds in \eqref{eq:second} as
\begin{eqnarray*}
\frac{ \theta \cdot u_{11} + u }{1-u} u_2 
    \le 10u + 165u^2 + 1421u^3 < 10u + 176u^2 \in \mathbb{F}
\end{eqnarray*}
and
\begin{equation}
    \frac{4u_s \cdot u_{16}}{1-u} < 5 u_s - \frac{1}{2}u_s, \quad 5u_s \in \mathbb{F}.
\label{eq:underflow}
\end{equation}
In \eqref{eq:underflow}, it is better to avoid the use of a subnormal number $u_s$ for the evaluation because of the inherent slowdown in CPU performance\footnote{The performance slowdown does not occur on a GPU}.

Finally, we have an upper bound for the right-hand side in \eqref{eq:th_bound}:
\begin{align}
    & \frac{u |\mathtt{fl}(\alpha_a + \alpha_b)| + |\delta_1'| + |\delta_2'| + |\delta_3'|}{1-u} \nonumber 
    < \frac{10u + 176u^2}{u_2} \gamma + \frac{5u_s\mathtt{fl}(\omega + 1)}{u_2} - \frac{1}{2}u_s \nonumber \\
    \le & \frac{\mathtt{fl}((10u + 176u^2)\gamma)}{u_1} + \frac{\mathtt{fl}(5u_s(\omega + 1))}{u_1}   
    \le \mathtt{fl}(\ (10u + 176u^2) \gamma  + 5u_s( \omega + 1))  \label{eq:filter01}
\end{align}
Algorithm~\ref{alg:proposed_filter01} is the floating-point filter based on \eqref{eq:filter01}.
\begin{algorithm}[H]
\caption{ICT($p_a, p_b, p_c, p_d$)}
\label{alg:proposed_filter01}
\begin{algorithmic}
\renewcommand{\algorithmicrequire}{\textbf{Input:}}
\Require 
Points($p_a(x_a, y_a)~,~p_b(x_b, y_b)~,~p_c(x_c, y_c)~,~p_d(x_d, y_d)$)
\State $adx \leftarrow x_a - x_d$, \quad $bdx \leftarrow x_b - x_d$, \quad $cdx \leftarrow x_c - x_d$
\State $ady \leftarrow y_a - y_d$, \quad $bdy \leftarrow y_b - y_d$, \quad $cdy \leftarrow y_c - y_d$
\State $\alpha_{a1} \leftarrow adx^2+ady^2$, \quad $\alpha_{a2} \leftarrow bdx*cdy-bdy*cdx$, \quad $\alpha_{a2'} \leftarrow |bdx*cdy|+|bdy*cdx|$
\State $\alpha_{a} \leftarrow \alpha_{a1}*\alpha_{a2}$
\State $\alpha_{a'} \leftarrow \alpha_{a1}*\alpha_{a2'}$
\State $\alpha_{b1} \leftarrow bdx^2+bdy^2$, \quad $\alpha_{b2} \leftarrow cdx*ady-cdy*adx$, \quad $\alpha_{b2'} \leftarrow |cdx*ady|+|cdy*adx|$
\State $\alpha_{b} \leftarrow \alpha_{b1}*\alpha_{b2}$
\State $\alpha_{b'} \leftarrow \alpha_{b1}*\alpha_{b2'}$
\State $\alpha_{c1} \leftarrow cdx^2+cdy^2$, \quad $\alpha_{c2} \leftarrow adx*bdy-ady*bdx$, \quad $\alpha_{c2'} \leftarrow |adx*bdy|+|ady*bdx|$
\State $\alpha_{c} \leftarrow \alpha_{c1}*\alpha_{c2}$
\State $\alpha_{c'} \leftarrow \alpha_{c1}*\alpha_{c2'}$
\State $det \leftarrow \alpha_{a}+\alpha_{b}+\alpha_{c}$
\State $\beta_{a} \leftarrow \alpha_{a1}*\alpha_{a2'}$
\State $\beta_{b} \leftarrow \alpha_{b1}*\alpha_{b2'}$
\State $\beta_{c} \leftarrow \alpha_{c1}*\alpha_{c2'}$
\State $errbound \leftarrow (10*u+176*u^2)*(\beta_{a}+\beta_{b}+\beta{c}) + 5*u_s*((\alpha_{a2'} + \alpha_{a1}) + (\alpha_{b2'}+ \alpha_{b1} ) + (\alpha_{c2'} + \alpha_{c1} ) + 1)$
\If{$|det|>errbound$}
\Return $det$
\Else
\State rationally calculate $det$ with GMP
\Return $det$
\EndIf
\end{algorithmic}
\end{algorithm}
If overflow occurs in Algorithm~\ref{alg:proposed_filter01}, 
$|det| > errbound$ is not satisfied, as described in \cite{ozaki-filter}.

\newpage
\section*{Acknowledgments}
This work was supported by JSPS KAKENHI Grant Number 23H03410 and 23K13020.
We would also like to acknowledge the support from the Kayamori Foundation of Informational Science Advancement under Grant Number K33-Ken-XXVI-596.
We thank Stuart Jenkinson, PhD, from Edanz for his help in  enhancing the English language quality of this manuscript.

\bibliographystyle{elsarticle-num}
\bibliography{vt}
\end{document}